\documentclass[manuscript]{aa}
\usepackage{natbib}
\bibpunct{(}{)}{;}{a}{}{,} 
\usepackage{graphicx}
\usepackage[varg]{txfonts}
\usepackage{color}

\def\note #1]{{\bf #1]}}

\def\dd{{\rm d}}

\newcommand{\msun}{\,\mathrm{M}_\odot}
\newcommand{\rsun}{\,\mathrm{R}_\odot}

\def\rbcz{r_{\rm bcz}}

\def\CA{{\cal A}}
\def\CB{{\cal B}}
\def\CI{{\cal I}}
\def\CG{{\cal G}}
\def\muHz{\,\mu{\rm Hz}}
\def\bolddelta{\delta\kern-0.45em\delta\kern-0.45em\delta}
\def\boldr{\mbox{\boldmath$r$}}

\def\bolddelr{\bolddelta \boldr}

\begin{document}

\title{The Aarhus red giants challenge II}
\subtitle{Stellar oscillations in the red giant branch phase}
\author{
	J.~Christensen-Dalsgaard\inst{\ref{instsac},\ref{instkitp}} \and V.~Silva Aguirre\inst{\ref{instsac}} \and S.~Cassisi\inst{\ref{instinaf},\ref{instinaf2}} \and M.~Miller~Bertolami\inst{\ref{instLP1},\ref{instLP2},\ref{instmpa}} \and A.~Serenelli\inst{\ref{instald},\ref{instbar}} \and   D.~Stello\inst{\ref{instsyd2},\ref{instsyd}, \ref{instsac}} \and A.~Weiss\inst{\ref{instmpa}} \and
G.~Angelou\inst{\ref{instmpa},\ref{instmps}} \and C.~Jiang\inst{\ref{instchi}} \and Y.~Lebreton\inst{\ref{instleb1},\ref{instleb2}} \and F.~Spada\inst{\ref{instmps}} \and
E.~P.~Bellinger\inst{\ref{instsac},\ref{instmps}} \and S.~Deheuvels\inst{\ref{instdeh}} \and R.~M.~Ouazzani\inst{\ref{instleb1},\ref{instsac}} \and 
A.~Pietrinferni\inst{\ref{instinaf}} \and  J.~R.~Mosumgaard\inst{\ref{instsac}} \and R.~H.~D.~Townsend\inst{\ref{instrich}} \and
T.~Battich\inst{\ref{instLP1},\ref{instLP2}} \and D.~Bossini\inst{\ref{instpor},\ref{instbir},\ref{instsac}} \and T. Constantino\inst{\ref{instexe}} \and P.~Eggenberger\inst{\ref{instegg}} \and S.~Hekker\inst{\ref{instmps},\ref{instsac}} \and A.~Mazumdar\inst{\ref{instanw}} \and A.~Miglio\inst{\ref{instbir},\ref{instsac}} \and K.~B.~Nielsen\inst{\ref{instsac}} \and M.~Salaris\inst{\ref{instsal}}
}
\institute{
Stellar Astrophysics Centre, Department of Physics and Astronomy, Aarhus University, Ny Munkegade 120, DK-8000 Aarhus C, Denmark\label{instsac} \and Kavli Institute for Theoretical Physics, University of California Santa Barbara, CA 93106-4030, USA\label{instkitp} \and
INAF-Astronomical Observatory of Abruzzo, Via M. Maggini sn, I-64100 Teramo, Italy\label{instinaf} \and
INFN - Sezione di Pisa, Largo Pontecorvo 3, 56127 Pisa, Italy\label{instinaf2} \and
Instituto de Astrof\'isica de La Plata, UNLP-CONICET, La Plata, Paseo del Bosque s/n, B1900FWA, Argentina\label{instLP1}\and
Facultad de Ciencias Astron\'omicas y Geof\'isicas, UNLP, La Plata, Paseo del Bosque s/n, B1900FWA, Argentina\label{instLP2}\and
Max-Planck-Institut f\"{u}r Astrophysics, Karl Schwarzschild Strasse 1, 85748, Garching, Germany\label{instmpa} \and
Instituto de Ciencias del Espacio (ICE-CSIC/IEEC), Campus UAB, Carrer de Can Magrans, s/n, 08193 Cerdanyola del Valles, Spain\label{instald} \and
Institut d'Estudis Espacials de Catalunya (IEEC), Gran Capita 4, E-08034, Barcelona, Spain\label{instbar} \and
Max-Planck-Institut f\"{u}r Sonnensystemforschung, Justus-von-Liebig-Weg 3, 37077, G\"{o}ttingen, Germany\label{instmps} \and
School of Physics and Astronomy, Sun Yat-Sen University, Guangzhou, 510275, China \label{instchi} \and
LESIA, Observatoire de Paris, PSL Research University, CNRS, Sorbonne Universit\'e, Univ. Paris Diderot, Sorbonne Paris Cit\'e, Meudon 92195, France\label{instleb1} \and
Univ Rennes, CNRS, IPR (Institut de Physique de Rennes) - UMR 6251, F-35000 Rennes, France\label{instleb2} \and
School of Physics, University of New South Wales, NSW, 2052, Australia\label{instsyd2} \and
Sydney Institute for Astronomy, School of Physics, University of Sydney, NSW 2006, Australia\label{instsyd} \and
IRAP, Universit\'e de Toulouse, CNRS, CNES, UPS, Toulouse, France\label{instdeh} \and
Department of Astronomy, 2535 Sterling Hall 475 N. Charter Street, Madison, WI 53706-1582, USA\label{instrich} \and
Instituto de Astrof\'isica e Ci\^encias do Espa\c co, Universidade do Porto, CAUP, Rua das Estrelas, PT-4150-762 Porto, Portugal\label{instpor} \and
School of Physics and Astronomy, University of Birmingham, Birmingham, B15 2TT, UK\label{instbir} \and
Physics and Astronomy, University of Exeter, Exeter, EX4 4QL, United Kingdom\label{instexe} \and
Observatoire de Gen\`eve, Universit\'e de Gen\`eve, 51 Ch. des Maillettes, CH-1290 Sauverny, Suisse\label{instegg} \and
Homi Bhabha Centre for Science Education, TIFR, V. N. Purav Marg, Mankhurd, Mumbai 400088, India\label{instanw} \and
Astrophysics Research Institute, Liverpool John Moores University, 146 Brownlow Hill, Liverpool L3 5RF, UK\label{instsal}
}

\abstract
{The large quantity of high-quality asteroseismic data that have been
obtained from space-based photometric missions and the accuracy of the 
resulting frequencies motivate a careful consideration of the accuracy 
of computed oscillation frequencies of stellar models, when applied 
as diagnostics of the model properties.}
{Based on models of red-giant stars that have been independently calculated
using different stellar evolution codes,
we investigate the
extent to which the differences in the model calculation affect the
model oscillation frequencies and other asteroseismic diagnostics.}
{For each of the models, which cover four different masses and different
evolution stages on the red-giant branch, we computed full sets of 
low-degree oscillation frequencies using a single pulsation code and,
from these frequencies, typical asteroseismic diagnostics.
In addition, we carried out preliminary analyses to relate differences
in the oscillation properties to the corresponding model differences.}
{In general, the differences in asteroseismic properties between the different
models greatly exceed the observational precision of these properties.
This is particularly true for the nonradial modes whose mixed acoustic 
and gravity-wave character makes them sensitive to the structure of
the deep stellar interior and, hence, to details of their evolution.
In some cases, identifying these differences led to improvements in the
final models presented here and in Paper~I; 
here we illustrate particular examples of this.
}
{Further improvements in stellar modelling are required  
in order fully to utilise
the observational accuracy to probe intrinsic limitations in the
modelling and improve our understanding of stellar internal physics.
However, our analysis of the frequency differences and their
relation to stellar internal properties provides a striking illustration
of the potential, in particular, of the mixed modes of red-giant stars
for the diagnostics of stellar interiors.}

\maketitle

\titlerunning{The Aarhus red giants challenge II}
\authorrunning{J. Christensen-Dalsgaard et al.}
\section{Introduction}\label{sec:int}


Space-based photometric observations of oscillations in red-giant stars
with the CoRoT \citep{Baglin2013}, {\it Kepler} \citep{Boruck2016} and,
since 2018, the TESS \citep{Ricker2014} missions
have provided a huge set of accurate oscillation frequencies and other 
properties for these stars.
These data provide the basis for detailed investigations of stellar structure
and evolution, as well as the application of stellar properties in other
areas of astrophysics, including the study of extra-solar planetary
systems and the structure and evolution of the Galaxy.
A necessary component of almost any analysis of such asteroseismic data
is the use of modelling of stellar structure and evolution and
the computation of oscillation frequencies for the resulting models.
Given the complexity of stellar modelling, it is a non-trivial task to
secure the required numerical and physical accuracy. 
Specifically, a full utilisation of the analysis of the
observations requires that the numerical errors in the computed 
properties are substantially smaller than the uncertainties
in the observations.
Although adequate convergence of the computations can, to some extent,
be tested by comparing results obtained with different numbers of meshpoints
or timesteps in the models,
more subtle errors in the calculations can probably only
be uncovered through comparisons of the results of independent codes
under carefully controlled conditions.

Extensive comparisons of this nature were organised for main-sequence stars
in connection with the CoRoT project \citep{Lebret2008}.
Detailed comparisons between stellar models for the Red Giant Branch stage
available in the literature have been discussed by \citet{Cassis1998},
\citet{Salari2002}, and \citet{Cassis2017}.
In the Aarhus Red Giant Challenge, we have so far concentrated on the numerical
properties of the computation of the stellar models.
Thus, the models are computed using, to the extent possible, 
the same input physics and basic parameters,
and the comparisons are carried out at carefully specified stages in the
evolution along the red-giant branch.
Differences between the model properties, including their oscillation
frequencies, should therefore reflect differences (and errors) in the numerical 
implementation of the solution of the equations of stellar evolution,
or in the implementation of the physics.
Also, we considered the effects of the 
resulting model differences on the computed oscillation frequencies,
hence providing a link to the asteroseismic observations,
with the goal of strengthening the basis for the analysis of the results of
space-based photometry.

The initial analysis has focused on models up to and including
the red-giant branch, emphasising the latter stage where energy production
takes place in a hydrogen-burning shell around an inert helium core.
\citet[][Paper~I]{Silva2020} presented model calculations
for selected models with masses of $1$, $1.5$, $2$, and $2.5 \msun$ on the
main sequence and the red-giant branch; 
the models analysed in detail were characterised in terms of radii chosen
such that the models are of interest in connection with the asteroseismic
investigations.
The calculations used nine different stellar-evolution codes;
Paper~I discusses the differences between the results
in terms of the overall properties of the models.
The present paper considers oscillation calculations,
using a single oscillation code,
for the models presented in Paper~I;
this includes some discussion of the relation between
the stellar structure and oscillation properties. 
A striking result is that the oscillation properties, in accordance
with the
potential for asteroseismic analyses, serve as a `magnifying glass' on
the differences in the stellar models, highlighting aspects where
different codes yield results that are significantly different at the
accuracy of the asteroseismic observations.

Further papers in this series will extend the analysis to
the so-called clump (or horizontal-branch) stars where, in addition
to the hydrogen-burning shell, there is helium fusion in the core;
this leads to a rather complex structure
and pulsation properties of the stars,
with interesting consequences for the comparison between models and
observed oscillations.
In addition, we shall consider comparisons between models computed with
`free physics', where each modeller chooses the parameters and physical
properties that would typically be used in the analysis of, for example,
{\it Kepler} data.
Finally, since the computation of stellar oscillations for these
evolved models involves a number of challenges, 
an additional consideration of the
comparisons between independent pulsation codes,
for a number of representative models, is also planned.

\section{Properties of red-giant oscillations}
\label{sec:oscprop}

\subsection{General properties}
\label{sec:genprop}


We consider oscillations of small amplitude and neglect effects of rotation and
other departures from spherical symmetry.
Then the modes depend on colatitude $\theta$ and longitude $\phi$ 
as spherical harmonics, $Y_l^m(\theta, \phi)$.
Here the degree $l$ measures the total number of nodes on the stellar surface
and the azimuthal order $m$ defines the number of nodal lines crossing the
equator.
Frequencies of spherically symmetric stars are independent of $m$.
In addition, a mode is characterised by the number and properties of the
nodes in the radial direction, which define the radial order $n$.
For reviews of the properties of stellar oscillations see, for example,
\citet{Aerts2010, Chapli2013, Hekker2017};
we discuss problems with the definition of the radial order
in Section \ref{sec:diporder}.

\begin{figure}[ht]
\includegraphics[angle=0,scale=0.75]{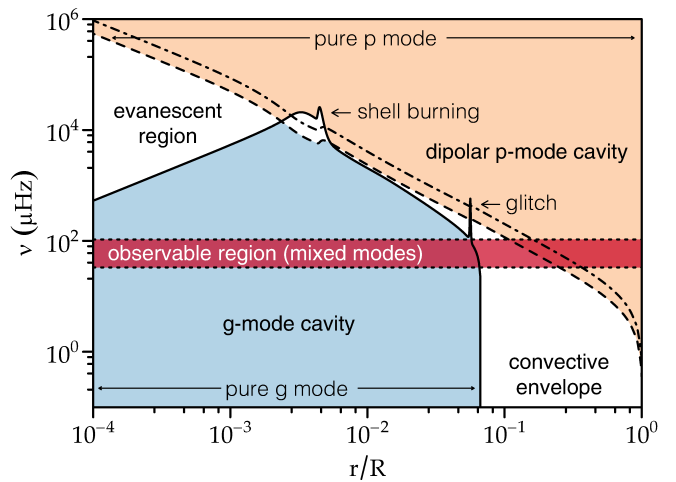}
\caption{
Characteristic frequencies $S_l/2 \pi$ for $l = 1$ and $2$ (dashed and
dot-dashed; cf.\ Eq.~\ref{eq:sl}) and 
$N/2 \pi$ (solid; cf.\ Eq.~\ref{eq:buoy}) in the {\tt ASTEC}
$1 \msun, 7 \rsun$ model.
The horizontal red band marks the region around $\nu_{\rm max}$,
the estimated frequency of maximum oscillation power
(cf.\ Eq.~\ref{eq:numaxscale}), where observed modes are expected.
The blue and orange areas indicate the corresponding regions of
g- and p-mode behaviour, for $l = 1$.
The glitch in the buoyancy frequency is caused by the near-discontinuity
in the hydrogen abundance resulting from the penetration, during
the first dredge-up, of the convective envelope into a region 
where the composition has been modified by nuclear reactions
(see also Paper~I).
}
\label{fig:charfreq}
\end{figure}

Radial modes, with $l = 0$, are purely acoustic, that is, standing sound waves.
Modes with $l > 0$ in red giants all have a mixed character, behaving 
as acoustic modes in the outer parts of the star and as internal gravity
waves in the core.
This is controlled by two characteristic frequencies of the star:
the acoustic (or Lamb) frequency
\begin{equation}
	S_l = {L c \over r} \; ,
\label{eq:sl}
\end{equation}
and the buoyancy (or Brunt-V\"ais\"al\"a) frequency $N$, given by
\begin{equation}
N^2 = g \left( {1 \over \Gamma_1} {\dd \ln p \over \dd r}
- {\dd \ln \rho \over \dd r} \right) \; .
\label{eq:buoy}
\end{equation}
Here $L = \sqrt{l(l+1)}$, $c$ is adiabatic sound speed, $r$ is distance to 
the centre, $g$ is local gravitational acceleration, $p$ is pressure,
$\rho$ is density, 
and $\Gamma_1 = (\partial \ln p / \partial \ln \rho)_{\rm ad}$ is 
adiabatic compressibility, the derivative being at constant specific entropy.
These frequencies are illustrated in Fig.~\ref{fig:charfreq} for 
a $1 \msun, 7 \rsun$ model, together with the typical observed
frequency range around the estimated frequency 
$\nu_{\rm max}$ at maximum oscillation power
\citep[see below; also][]{Hekker2017}.
In the outer region, where $\nu > S_l$ and $\nu > N$ 
(i.e.\ the p-mode cavity), 
the mode behaves acoustically,
while in the core where $\nu < S_l$ and $\nu < N$ 
(i.e.\ the g-mode cavity) 
the mode behaves like an internal gravity wave.
In the intermediate region, the mode has an exponential behaviour; the
extent of this so-called evanescent region controls the coupling between
the acoustic and gravity-wave behaviour in the given mode.
As a result of these properties, all nonradial modes, with $l > 0$,
have a mixed nature, with sensitivity both to the outer layers and to the
core.
For detailed discussions of such mixed modes see, for example,
\citet{Hekker2017, Mosser2018}; and references therein.
Here we present some of the properties of the modes which are useful
for the following analysis.

Acoustic modes of low degree have the following asymptotic behaviour
\citep{Shibah1979, Tassou1980, Gough1993}:
\begin{equation}
\nu_{n_{\rm p}l} \approx 
\Delta \nu \left(n_{\rm p} + {l \over 2} + \epsilon_{\rm p} \right)
+ d_{n_{\rm p}l} \; ,
\label{eq:pasymp}
\end{equation}
for the cyclic frequencies $\nu_{n_{\rm p}l}$.
Here the asymptotic expression for the large frequency separation
$\Delta \nu$ is
\begin{equation}
\Delta \nu = \Delta \nu_{\rm as} = 
\left( 2 \int_0^{R_*} {\dd r \over c} \right)^{-1} \; ,
\label{eq:dnu}
\end{equation}
that is, the inverse of twice the sound travel time between the centre and 
the so-called acoustic surface \citep{Houdek2007},
at a distance $R_*$ from the centre, in the stellar atmosphere.
In the strict asymptotic analysis, $\epsilon_{\rm p}$ is a constant
and the small higher-order effects are contained in $d_{n_{\rm p}l}$
\citep[see also][]{Mosser2013}.
Here, however, we adopt the formalism of \citet{Roxbur2013} and regard
$\epsilon_{\rm p}$ as a phase function depending on frequency but not on
degree, determined by the properties of the near-surface layers
\citep[see also][]{Christ1992};
this allows us to assume that $d_{n_{\rm p}l}$ is 0 for $l = 0$.
For the purely acoustic radial modes, Eq.~(\ref{eq:pasymp}) provides 
an approximation to the frequencies as a function 
of mode order $n_{\rm p}$; 
for the mixed nonradial modes the acoustically dominated modes (known as
p-m modes) approximately satisfy the relation for an order $n_{\rm p}$
characterising the acoustic behaviour.

Observed and computed acoustic-mode frequencies 
follow Eq.~(\ref{eq:pasymp}) fairly closely, to leading order, 
although the value of the
large frequency separation obtained from Eq.~(\ref{eq:dnu}) is not 
sufficiently accurate to be applied to comparisons with observations.
In the analysis of observed frequencies, various techniques can be used
to determine the large frequency separation 
\citep[e.g.][]{Huber2009, Mosser2009, Hekker2010, Kallin2010}.
The relation between different measures of the large frequency separation
for stellar models
was discussed by \citet{Belkac2013} and \citet{Mosser2013}.
Here we follow \citet{White2011} and \citet{Mosser2013}
and consider $\Delta \nu_{\rm fit}$
obtained from a weighted least-squares fit 
to frequencies of radial modes around the frequency
$\nu_{\rm max}$ of maximum power (see Eq.~\ref{eq:numaxscale} below),
with a weight reflecting an estimate of the mode power.
Some details on the fit are provided in Section \ref{sec:dnufit}.
We do note, however, that this procedure does not fully
represent the weighting in the analyses of observational data,
which are typically done directly from the observed power spectrum,
for example through a cross-correlation analysis, without reference
to the individual mode frequencies.
Even so, obtaining $\Delta \nu_{\rm fit}$ from a fit to the computed frequencies
provides a convenient way to compare the results for different codes.

Modes dominated by internal gravity waves require density variations
over spherical surfaces in the star and are therefore only found for $l > 0$.
For such pure g modes, the periods $\Pi_{n_{\rm g}l} = 1 / \nu_{n_{\rm g}l}$
satisfy
\begin{equation}
\Pi_{n_{\rm g}l} = \Delta \Pi_l \left( n_{\rm g} + \epsilon_{\rm g} \right) 
\label{eq:gasymp}
\end{equation}
\citep[e.g.][]{Shibah1979, Tassou1980}, where
\begin{equation}
\Delta \Pi_l = {\Pi_0 \over L} \; , \qquad
\Pi_0 = 2 \pi^2 \left( \int N {\dd r \over r} \right)^{-1} \; ,
\label{eq:dpi}
\end{equation}
the integral being over the gravity-wave cavity, and $\epsilon_{\rm g}$ is
a phase, the so-called gravity offset, that may depend on $l$.
Mixed modes dominated by the gravity-wave behaviour (the g-m modes)
approximately satisfy Eq.~(\ref{eq:gasymp}), with $n_{\rm g}$ being
an order characterising the g-mode behaviour
(see also Fig.~\ref{fig:inertia} below),
and hence provide a measure of $\Delta \Pi_l$.
However, additional important characteristics are provided by the
measure $q$ of the coupling between the g- and p-mode cavities
and $\epsilon_{\rm g}$,
which provide information about the evanescent region
and the upper part of the g-mode cavity
\citep[e.g.][]{Takata2016, Hekker2017, Pincon2019}.

\subsection{Observational properties}
\label{sec:obsprop}

The information available from the observed frequencies of oscillation
depends strongly on the quality of the data.
The most visible modes are the
acoustically dominated (p-m) modes, which provide information about the
overall properties of the star.
They are characterised by the large frequency separation $\Delta \nu$
(cf.\ Eq.~\ref{eq:dnu}) and the frequency $\nu_{\rm max}$ of maximum power.
It follows from homology scaling that, approximately,
$\Delta \nu \propto \bar \rho^{1/2}$
where $\bar \rho$ is the mean density of the star.
Specifically,
\begin{equation}
\Delta \nu \simeq \left({M \over \msun}\right)^{1/2} 
\left( {R \over \rsun} \right)^{-3/2} \Delta \nu_\odot \; ,
\label{eq:dnuscale}
\end{equation}
valid for both $\Delta \nu_{\rm as}$ and $\Delta \nu_{\rm fit}$,
where $\Delta \nu_\odot$ is the corresponding value for the Sun.
A characteristic observed value is $\Delta \nu_\odot \simeq 135.1 \muHz$.
Also, observationally \citep{Brown1991} and with some theoretical support
\citep{Belkac2011} $\nu_{\rm max}$ scales as the acoustic cut-off frequency,
such that
\begin{equation}
\nu_{\rm max} \simeq {M \over \msun} \left( {R \over \rsun} \right)^{-2}
\left({T_{\rm eff} \over T_{\rm eff, \odot}} \right) ^{-1/2} 
\nu_{\rm max, \odot} \; ,
\label{eq:numaxscale}
\end{equation}
where $\nu_{\rm max, \odot} \simeq 3090 \muHz$ is the frequency at maximum
power for the Sun.
In an extensive analysis of a large sample of {\it Kepler} red giants,
\cite{Yu2018} found
relative uncertainties in the large frequency separation $\Delta \nu$ below
0.1\,\%, in some cases, with a median value of 0.6\,\%,
while the median uncertainty in $\nu_{\rm max}$ was 1.6\,\%.
From the scaling relations in Eqs.~(\ref{eq:dnuscale}) and
(\ref{eq:numaxscale}), stellar masses and radii
can be determined \citep[see, for example,][]{Kallin2010, Yu2018}.
In practice, departures from the strict scaling relations,
for example caused by departures from homology, need to be taken into account
\citep[see][for a review]{Hekker2019};
we return to this below in connection with a discussion of the scaling in 
Eq.~(\ref{eq:dnuscale}).

The g-m modes, with a large component of internal gravity wave,
provide strong constraints on stellar properties
\citep[for an early example, see][]{Hjorri2017}.
With improved analysis, further details on the g-m mode properties
are becoming available \citep{Mosser2018}, providing further constraints
on the stellar internal properties.
Uncertainties in individual frequencies for both acoustic and mixed modes
are as low as $0.01 \muHz$
\citep{Corsar2015, deMont2018}, corresponding to relative uncertainties of 
order $10^{-4}$.
Analysis of observed data on mixed modes in red giants has so
far mainly been carried out in terms of determinations of the asymptotic
properties characterised by $\Delta \Pi_l$, $q$ and $\epsilon_g$
as obtained from fitting a full asymptotic expression 
to the observed frequencies.
\citep[e.g.][]{Beddin2011, Mosser2014, Mosser2017, Mosser2018}.
Uncertainties in the dipolar period spacing 
$\Delta \Pi_1$ of around $0.1\,{\rm s}$ were quoted by
\citet{Hekker2018, Mosser2018}.
Detailed model fits to individual frequencies should also be feasible
but have so far apparently not seen much use.

\subsection{Oscillation properties of red-giant models}


\begin{figure}[ht]
\includegraphics[angle=0,scale=0.5]{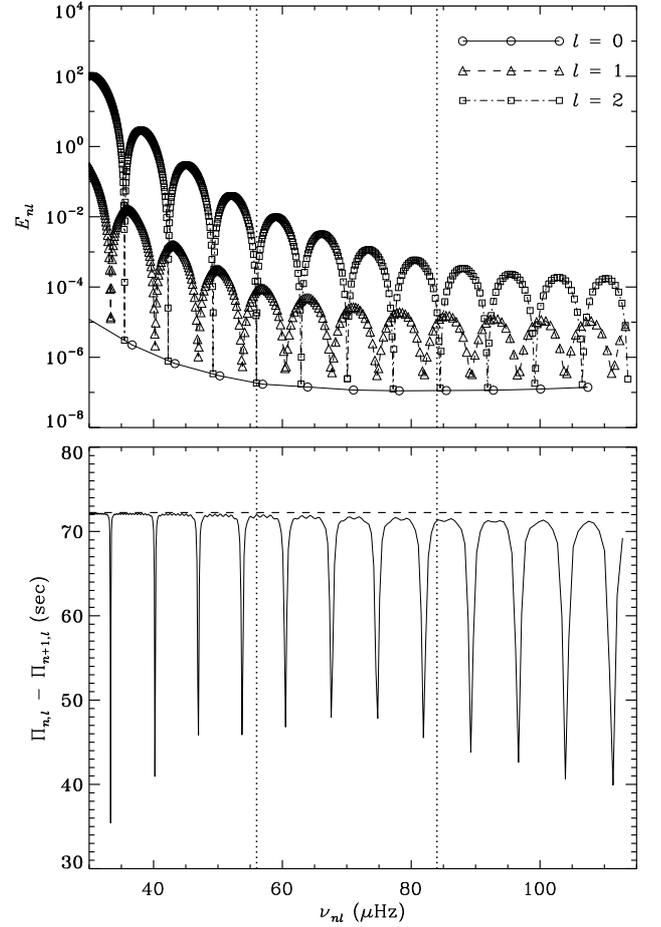}
\caption{
Top: Mode inertia (cf Eq.~\ref{eq:inertia}) for modes of degree $l = 0$
(solid line, circles), $1$ (dashed line, triangles), and $2$
(dot-dashed line, squares),
in the {\tt ASTEC} $1 \msun, 7 \rsun$ model.
Bottom: Separation between periods of adjacent modes with $l = 1$
in this model, plotted against frequency. 
The horizontal dashed line shows the asymptotic period spacing
$\Delta \Pi_1 = \Pi_0/\sqrt{2}$ (cf.\ Eqs.~\ref{eq:gasymp} and \ref{eq:dpi}).
The heavy vertical dotted lines show the frequency interval
where power is half its maximum
value, according to the fit of \citet{Mosser2012} 
(see also Section~\ref{sec:dnufit}).
}
\label{fig:inertia}
\end{figure}

To compare the oscillation properties of the models involved in the challenge
the equations of adiabatic oscillations were solved using the code
{\tt ADIPLS} \citep{Christ2008a}.
This code was compared in detail with other pulsation codes by 
\citet{Moya2008} and, more recently as part of the present project,
with the GYRE code \citep{Townse2013}.
Owing to the condensed core of red-giant stars
and the resulting very high value of the buoyancy frequency,
modes of very high radial order
are involved, requiring some care in the preparation of the models for
the oscillation calculations; 
some details of these procedures are discussed in Section \ref{sec:compproc};
in Section \ref{sec:numprec} we estimate the numerical errors 
in the resulting frequencies,
both the intrinsic errors of the oscillation calculation and the effects
on the frequencies
from the errors in the computation of the {\tt ASTEC} models, which are used
for reference in the comparisons.
Comparisons between models should be carried out at fixed mode order,
requiring a determination of the order of the computed modes.
For dipolar modes, this gives rise to some complications, compounded by
inconsistencies in the structure very near the centre in some models, 
as discussed in Section \ref{sec:diporder}.
Computed frequencies for all models, as well as the model structure,
are provided at the website of the project.%
\footnote{https://github.com/vsilvagui/aarhus\_RG\_challenge}

To characterise the properties of the modes a very useful quantity is
the normalised mode inertia,
\begin{equation}
E = {\int_V \rho |\bolddelr|^2 \dd V \over M |\bolddelr|^2_{\rm phot}} \; ,
\label{eq:inertia}
\end{equation}
where $\bolddelr$ is the displacement vector and `phot' indicates the
photospheric value,
defined at the location where the temperature equals the effective temperature;
the integral is over the volume $V$ of the star. 
In Fig.~\ref{fig:inertia} the top panel shows the inertia 
for a $1 \msun, 7 \rsun$ model computed with {\tt ASTEC}.
Predominantly acoustic (p-m) modes have their largest amplitude in the outer
layers of the star, where $\rho$ is small, and hence $E$ is relatively small,
while g-m modes have large inertias.
For the radial modes, the inertia decreases strongly with increasing frequency
at low frequency, while it is almost constant at higher frequency.
For $l = 1$ and $2$, there is evidently a very high density of modes,
most of which have inertia much higher than those of the radial modes
and hence are predominantly of g-m character.
However, there are clear acoustic resonances where the inertia approaches
the radial-mode values and the modes are predominantly of p-m character.
The frequencies of these resonances satisfy the asymptotic relation
in Eq.~(\ref{eq:pasymp}); 
in particular, $l = 0$ and $2$ modes separated by one in the acoustic
order $n_{\rm p}$ have frequencies at a small separation determined by
the term in $d_{n_{\rm p}l}$.
It should also be noticed that the minimum inertia at the resonances
is substantially lower for $l = 2$ than for $l = 1$:
as shown in Fig.~\ref{fig:charfreq} the evanescent region
is broader for $l = 2$, leading to a weaker coupling
and hence to a more dominant acoustic character of the mode at a resonance.

This mixed character of the modes is also visible in the bottom panel
of Fig.~\ref{fig:inertia},
which shows the period spacing between adjacent dipolar modes in
the same model.
For most of the modes, particularly at low frequency, 
the computed period spacing is very close
to the asymptotic value, indicated by the horizontal dashed line.
However, at the acoustic resonances where the modes take on a p-m 
character the period spacing is strongly reduced;
we note that these resonances take place at frequencies approximately
satisfying Eq.~(\ref{eq:pasymp}).

\section{Results of model comparisons}
\label{sec:res}

\subsection{Stellar models}

We computed oscillation properties of the models highlighted in Paper~I.
We note in particular that two sets of models have been considered.
In one (in the following the solar-calibrated models), the mixing-length 
parameter $\alpha_{\rm MLT}$ was adjusted in each code to achieve a 
photospheric radius of $6.95508 \times 10^{8} \, {\rm m}$
at the age of 4.57 Gyr of main-sequence evolution for the $1 \msun$ model.
In the second (in the following the RGB-calibrated models), $\alpha_{\rm MLT}$
was fixed for each track by requiring a specific effective temperature
$T_{\rm eff}$ for the $7 \rsun$ models on the $1$ and $1.5 \msun$ tracks
and the $10 \rsun$ models on the $2$ and $2.5 \msun$ tracks.
In the present section, we generally focus on the solar-calibrated models;
results for the RGB-calibrated models are provided in Appendix~\ref{sec:rgbcal}.

The following evolution codes were used:

\begin{itemize}

\item {\tt ASTEC}: the Aarhus STellar Evolution Code; see \citet{Christ2008b}.

\item {\tt BaSTI}: Bag of Stellar Tracks and Isochrones; see \citet{Pietri2013}.

\item {\tt CESAM2k}: Code d'Evolution Stellaire Adaptatif et Modulaire,
	2000 version; see \citet{Morel2008}.

\item {\tt GARSTEC}: the GARching STellar Evolution Code; see \citet{Weiss2008}.

\item {\tt LPCODE}: the La Plata stellar evolution Code; see \citet{Miller2016}.

\item {\tt MESA}: Modules for Experiments in Stellar Astrophysics, version
	6950; see \citet{Paxton2013}.

\item {\tt MONSTAR}: the Monash version of the Mt Stromlo evolution code;
	see \citet{Consta2015}.

\item {\tt YaPSI}: the Yale Rotational stellar Evolution Code, as used
	in the Yale-Potsdam Stellar Isochrones; see \citet{Spada2017}.

\item {\tt YREC}: the Yale Rotating stellar Evolution Code;
	see \citet{Demarq2008}.

\end{itemize}

We note that in order to avoid effects of different extents of
the atmosphere in models from different codes for a given set of parameters,
the models were truncated in the atmosphere at a height corresponding to
the code with the smallest atmospheric extent, for the given case.

Further details about the codes and the models are provided in Paper~I.

\begin{figure}[ht]
\includegraphics[angle=0,scale=0.5]{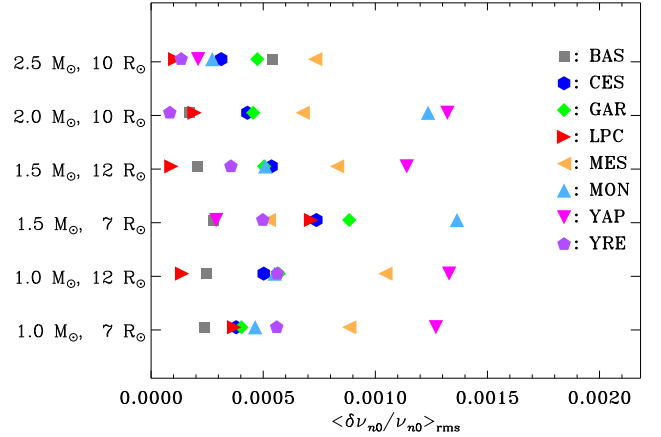}
\caption{
Root-mean-square relative differences,
in the solar-calibrated case,
in radial-mode frequencies relative to
the {\tt ASTEC} results, in the sense (model) - ({\tt ASTEC});
the different codes are identified by the symbol shape and colour 
and labelled by
the abbreviated name of the code: BAS ({\tt BaSTI}), CES ({\tt CESAM}),
GAR ({\tt GARSTEC}), LPC ({\tt LPCODE}),
MES ({\tt MESA}), MON ({\tt MONSTAR}), YAP ({\tt YaPSI}), and YRE ({\tt YREC}).
}
\label{fig:diffdr0_solar}
\end{figure}

\begin{table*}[ht]
\caption{Large frequency separations $\Delta \nu_{\rm fit}$ in $\muHz$
obtained from fits to radial-mode frequencies as functions
of mode order (cf.\ Eq.~\ref{eq:pasymp} and Section \ref{sec:dnufit})
for solar-calibrated models.
}
\label{tab:dnufrqsolar}
\centering
\begin{tabular}{r r c c c c c c c c c}
\hline\hline
$M/\msun$ & $R/\rsun$ & {\tt ASTEC} & {\tt BaSTI} & {\tt CESAM} & {\tt GARSTEC} & {\tt LPCODE} & {\tt MESA} & {\tt MONSTAR} & {\tt YAP} & {\tt YREC} \\
\hline
\smallskip
1.0 &  7.0 &  7.087 &  7.093 &  7.100 &  7.089 &  7.090 &  7.077 &  7.084 &  7.100 &  7.090\\
1.0 & 12.0 &  3.130 &  3.133 &  3.137 &  3.131 &  3.130 &  3.124 &  3.128 &  3.137 &  3.131\\
1.5 &  7.0 &  8.780 &  8.783 &  8.799 &  8.774 &  8.789 &  8.768 &  8.762 &  8.778 &  8.785\\
1.5 & 12.0 &  3.876 &  3.879 &  3.884 &  3.876 &  3.875 &  3.869 &  3.873 &  3.882 &  3.876\\
2.0 & 10.0 &  5.949 &  5.952 &  5.958 &  5.949 &  5.951 &  5.940 &  5.938 &  5.938 &  5.948\\
2.5 & 10.0 &  6.739 &  6.746 &  6.744 &  6.735 &  6.737 &  6.726 &  6.736 &  6.744 &  6.738\\
\hline
\end{tabular}
\end{table*}

\subsection{Acoustic properties}
\label{sec:acoustic}

We first consider the properties of the acoustically-dominated
oscillations, as characterised by the radial modes.
As an indication of the frequency differences between different codes,
Fig.~\ref{fig:diffdr0_solar}
shows the root-mean-square relative differences
in radial-mode frequencies between the various codes and {\tt ASTEC},
including all modes up to the acoustic cut-off frequency.
We note that they are far bigger than the
observational uncertainties of the individual frequencies
(cf.\ Section~\ref{sec:obsprop}).

\begin{figure}[ht]
\includegraphics[angle=0,scale=0.5]{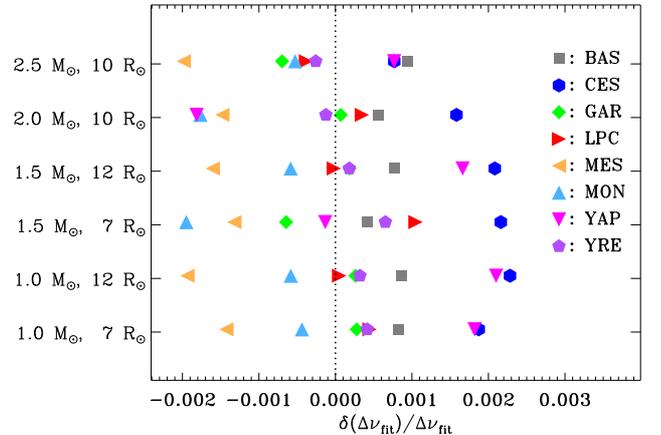}
\caption{
Relative differences for solar-calibrated models
in the large frequency separations
$\Delta \nu_{\rm fit}$ obtained from fits to the radial-mode 
frequencies as functions of mode order (cf.\ Section \ref{sec:dnufit}),
compared with the {\tt ASTEC} results, in the sense (model) -- ({\tt ASTEC});
the different codes are identified by the symbol shape and colour 
(cf.\ caption to Fig.~\ref{fig:diffdr0_solar}).
\label{fig:diffdnufrqsolar}
}
\end{figure}

The large frequency separation between acoustic modes is an important
asteroseismic diagnostics.
As discussed in Section~\ref{sec:genprop}, we characterise the observable values
by the result $\Delta \nu_{\rm fit}$ of fitting the computed radial-mode
frequencies to Eq.~(\ref{eq:pasymp}),
representing $\epsilon_{\rm p}$ by a quadratic expression
in mode order (cf.\ Section~\ref{sec:dnufit}).
Table~\ref{tab:dnufrqsolar} provides values obtained 
from these fits in the solar-calibrated case, 
and Fig.~\ref{fig:diffdnufrqsolar} shows relative differences
for these fitted values, relative to the ASTEC models.
In most cases the relative differences are below 0.2\,\%,
comparable with or somewhat bigger than the
observational uncertainties of around 0.1\,\%
(cf.\ Section~\ref{sec:obsprop}).

\begin{table*}
\caption{Correction factors $f_{\Delta \nu}$ (cf.\ Eq.~\ref{eq:dnucalscale})
between the large frequency separation $\Delta \nu_{\rm fit}$ obtained from 
a fit to radial-mode frequencies and the value obtained
from homology scaling for solar-calibrated models.
}
\label{tab:dnusclsolar}
\centering
\begin{tabular}{r r c c c c c c c c c}
\hline\hline
$M/\msun$ & $R/\rsun$ & {\tt ASTEC} & {\tt BaSTI} & {\tt CESAM} & {\tt GARSTEC} & {\tt LPCODE} & {\tt MESA} & {\tt MONSTAR} & {\tt YAP} & {\tt YREC} \\
\hline
\smallskip
1.0 &  7.0 &  0.9655 &  0.9661 &  0.9667 &  0.9658 &  0.9660 &  0.9656 &  0.9656 &  0.9682 &  0.9658\\
1.0 & 12.0 &  0.9572 &  0.9578 &  0.9588 &  0.9575 &  0.9573 &  0.9568 &  0.9571 &  0.9601 &  0.9573\\
1.5 &  7.0 &  0.9767 &  0.9769 &  0.9782 &  0.9761 &  0.9777 &  0.9768 &  0.9752 &  0.9774 &  0.9771\\
1.5 & 12.0 &  0.9677 &  0.9682 &  0.9691 &  0.9679 &  0.9677 &  0.9676 &  0.9676 &  0.9702 &  0.9677\\
2.0 & 10.0 &  0.9786 &  0.9789 &  0.9795 &  0.9787 &  0.9789 &  0.9786 &  0.9773 &  0.9777 &  0.9783\\
2.5 & 10.0 &  0.9915 &  0.9923 &  0.9917 &  0.9909 &  0.9912 &  0.9911 &  0.9915 &  0.9932 &  0.9911\\
\hline
\end{tabular}
\end{table*}

\begin{figure}[ht]
\includegraphics[angle=0,scale=0.5]{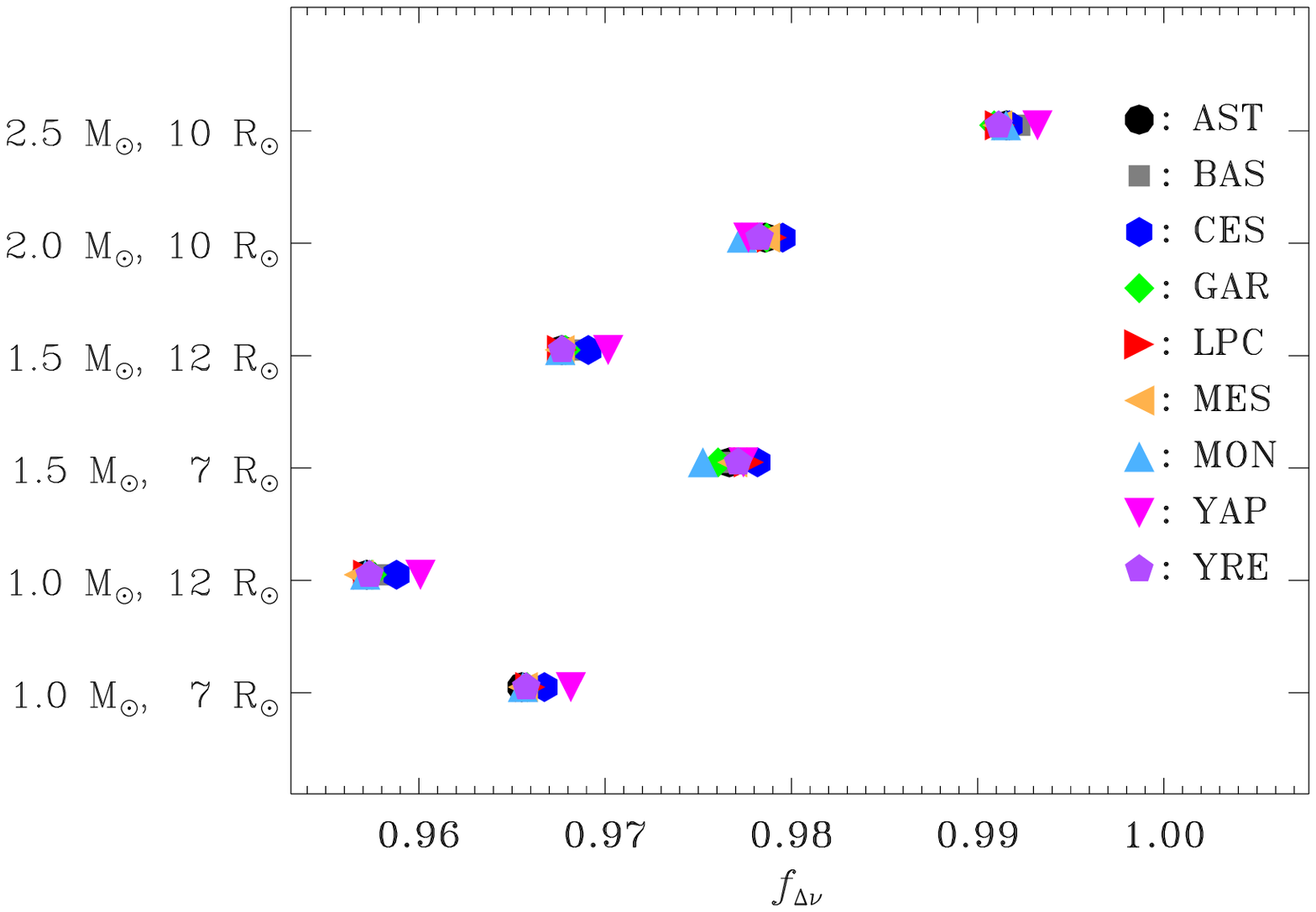}
\caption{
Correction factor $f_{\Delta \nu}$,
for solar-calibrated models,
in the scaling relation for the
large frequency separation $\Delta \nu_{\rm fit}$ obtained from
fits to the radial-mode frequencies (cf.\ Eq.~\ref{eq:dnucalscale}).
The different codes are identified by the symbol shape and colour 
(cf.\ caption to Fig.~\ref{fig:diffdr0_solar}), with the addition of AST
(for {\tt ASTEC}).
}
\label{fig:dnusclsolar}
\end{figure}

\begin{figure}[htpb]
\includegraphics[angle=0,scale=0.5]{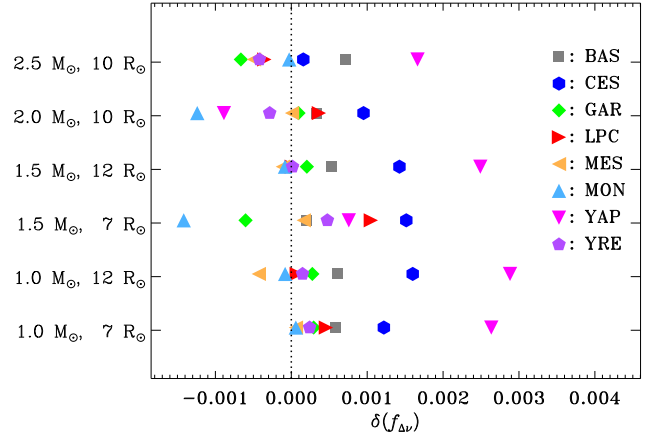}
\caption{
Differences relative to the {\tt ASTEC} model in the solar-calibrated case,
in the sense (model) -- ({\tt ASTEC}),
in the correction factors $f_{\Delta \nu}$ (cf.\ Eq.~\ref{eq:dnucalscale})
relating the large frequency separation $\Delta \nu_{\rm fit}$ obtained from 
a fit to radial-mode frequencies and the value obtained
from homology scaling.
The different codes are identified by the symbol shape and colour 
(cf.\ caption to Fig.~\ref{fig:diffdr0_solar}).
\label{fig:diffdnusclsolar}
}
\end{figure}

The scaling relation in Eq.~(\ref{eq:dnuscale}) is fundamental in the analysis
of global seismic observations, but the underlying assumed homology scaling 
is not exact.
Thus, it is often corrected by including a factor $f_{\Delta \nu}$
on the right-hand side \citep[e.g.][]{White2011, Rodrig2017};
\citep[see][for an application]{Sharma2016}.
Within the present model analysis we replace Eq.~(\ref{eq:dnuscale}) by
\begin{equation}
\Delta \nu_{\rm fit} = f_{\Delta \nu} \left({M \over \msun}\right)^{1/2} 
\left( {R \over \rsun} \right)^{-3/2} \Delta \nu_{\rm fit}^{\rm (cal)} \; ,
\label{eq:dnucalscale}
\end{equation}
where $\Delta \nu_{\rm fit}^{\rm (cal)}$ is the large separation resulting
from a fit to the radial modes of the $(1 \msun, 1 \rsun)$
models used to calibrate the mixing length in the solar-calibrated case.
The resulting values of the correction factor $f_{\Delta \nu}$ 
are shown in Table~\ref{tab:dnusclsolar} and
Fig.~\ref{fig:dnusclsolar} for the solar-calibrated case.
The dominant variation is that $f_{\Delta \nu}$ approaches unity for 
the most massive model, in accordance with the results obtained by
\citet{Guggen2017, Rodrig2017}.
Differences in $f_{\Delta \nu}$ between the different
codes relative to ASTEC are shown in Fig.~\ref{fig:diffdnusclsolar}.
We note that for any given model case there is a spread of 
around $\pm 0.002$ between the values
of $f_{\Delta\nu}$ obtained by the different evolution codes.
As pointed out by, for example, \citet{Sharma2016} the radius and mass
obtained from direct scaling analysis of global asteroseismic observables
scale as, respectively, $f_{\Delta\nu}^{-2}$ and $f_{\Delta\nu}^{-4}$.
Thus, the spread between the codes would correspond to variations of around
0.4 and 0.8\,\% in the inferred radii and masses,
when using Eq.~(\ref{eq:dnucalscale}) to analyse observed data.

\begin{table*}[ht]
\caption{Frequency $\nu_{\rm max}$, in $\muHz$, of maximum oscillation power
estimated from Eq.~(\ref{eq:numaxscale}) for solar-calibrated models.}
\label{tab:numaxsolar}
\centering
\begin{tabular}{r r c c c c c c c c c}
\hline\hline
$M/\msun$ & $R/\rsun$ & {\tt ASTEC} & {\tt BaSTI} & {\tt CESAM} & {\tt GARSTEC} & {\tt LPCODE} & {\tt MESA} & {\tt MONSTAR} & {\tt YAP} & {\tt YREC} \\
\hline
\smallskip
1.0 &  7.0 &  69.76 &  69.97 &  70.03 &  69.93 &  69.74 &  69.61 &  69.90 &  69.87 &  69.78\\
1.0 & 12.0 &  24.29 &  24.36 &  24.38 &  24.36 &  24.29 &  24.24 &  24.34 &  24.33 &  24.30\\
1.5 &  7.0 & 102.57 & 102.85 & 102.91 & 102.82 & 102.55 & 102.35 & 102.80 & 102.73 & 102.61\\
1.5 & 12.0 &  35.76 &  35.86 &  35.88 &  35.85 &  35.75 &  35.69 &  35.83 &  35.82 &  35.77\\
2.0 & 10.0 &  67.05 &  67.24 &  67.27 &  67.24 &  67.06 &  66.93 &  67.21 &  67.17 &  67.06\\
2.5 & 10.0 &  82.64 &  82.85 &  82.92 &  82.88 &  82.65 &  82.49 &  82.82 &  82.79 &  82.65\\
\hline
\end{tabular}
\end{table*}

\begin{figure*}[ht]
\includegraphics[angle=0,scale=0.5]{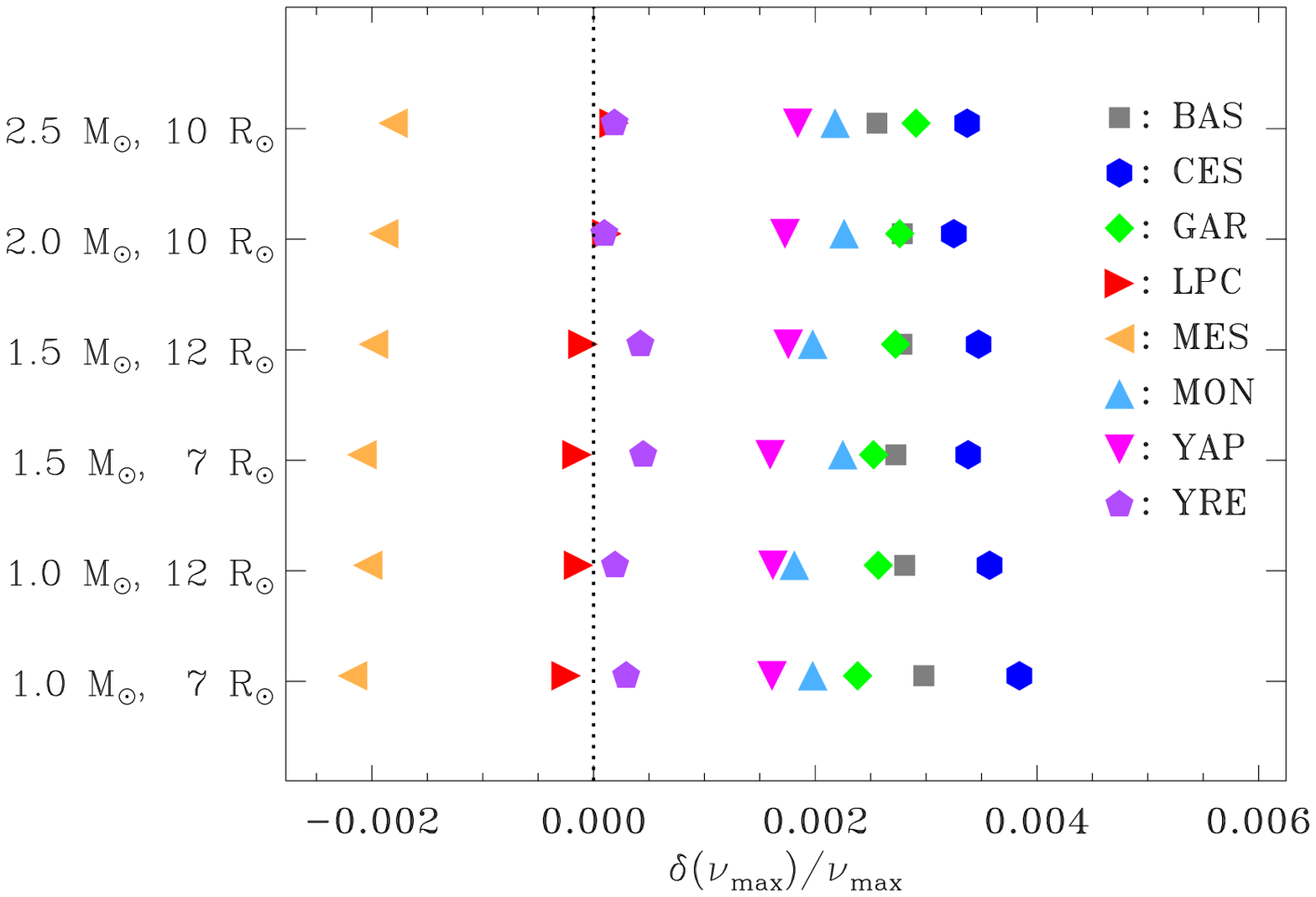}
\includegraphics[angle=0,scale=0.5]{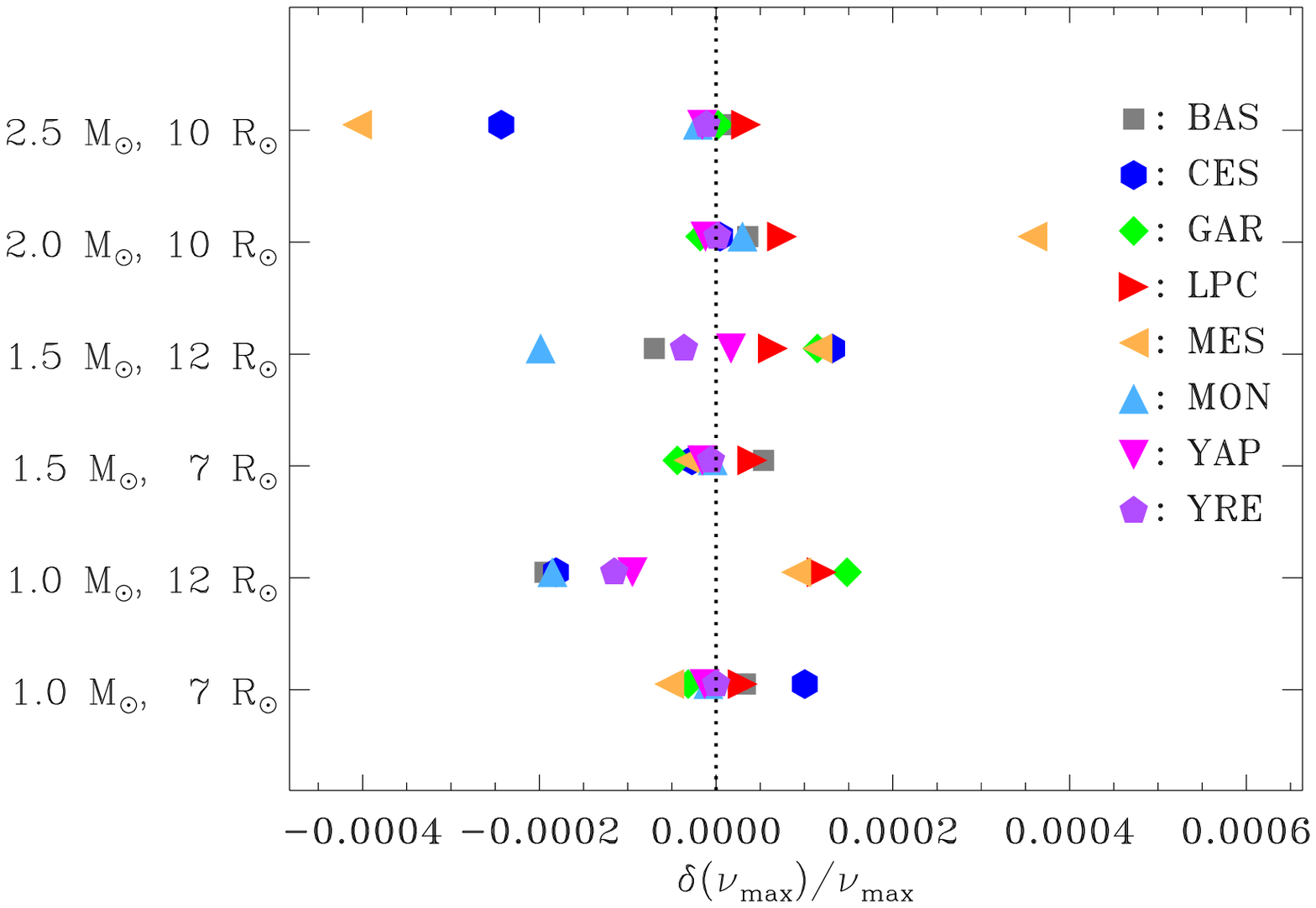}
\caption{
Relative differences in the estimated frequency $\nu_{\rm max}$
of maximum power,
compared with the {\tt ASTEC} results, in the sense (model) - ({\tt ASTEC});
the different codes are identified by the symbol shape and colour 
(cf.\ caption to Fig.~\ref{fig:diffdr0_solar}).
The left panel shows results for the solar-calibrated models,
and the right panel for the RGB-calibrated models (note the different
scales on the abscissas).
}
\label{fig:diffnumax}
\end{figure*}

The frequency $\nu_{\rm max}$ of maximum power plays 
an important role for asteroseismic inference.
In view of this,
we include a brief analysis of the differences in $\nu_{\rm max}$
between the models,
even though these differences essentially reflect the differences
in $T_{\rm eff}$, already discussed in Paper~I,
given that the comparison is carried out at fixed target model radius,
and with a constraint on $GM/R^3$.
Table~\ref{tab:numaxsolar} shows values of $\nu_{\rm max}$,
estimated from Eq.~(\ref{eq:numaxscale}), for solar-calibrated models,
while Fig.~\ref{fig:diffnumax} shows
relative differences in $\nu_{\rm max}$, compared with the {\tt ASTEC} values
for both solar- and RGB-calibrated models.
For the solar-calibrated models, there is a spread of order 50\,K between
different codes for any given model, corresponding to roughly 1\,\%
(cf.\ Figs.~2 and 7, and Tables C.2 -- C.4, of Paper~I;
see also Fig.~\ref{fig:diffteff} below).
For the RGB-calibrated models,
where $T_{\rm eff}$ was explicitly constrained on the RGB,
the spread is less than 4\,K, or 0.1\,\%.
Thus, the differences are smaller by about an order of magnitude
for the RGB-calibrated models.
We note, however, that in either case the differences 
in $\nu_{\rm max}$ are far smaller than
the typical observational uncertainty of 1.6\,\% \citep{Yu2018}
in this quantity.

Although the asymptotic value, $\Delta \nu_{\rm as}$, of the
large frequency separation (cf.\ Eq.~\ref{eq:dnu}) does not
provide sufficient accuracy for comparison with observed frequencies,
it is still an interesting diagnostics of the acoustic properties of the
models.
We analyse it in Appendix~\ref{sec:dnuprop}.

\subsection{Mixed modes}
\label{sec:mixmodes}

\begin{table*}[ht]
\caption{Asymptotic dipolar g-mode period spacings $\Delta \Pi_1$ in s 
(cf.\ Eq.~\ref{eq:dpi}) for solar-calibrated models.
}
\label{tab:perspacsolar}
\centering
\begin{tabular}{r r r r r r r r r r r}
\hline\hline
$M/\msun$ & $R/\rsun$ & {\tt ASTEC} & {\tt BaSTI} & {\tt CESAM} & {\tt GARSTEC} & {\tt LPCODE} & {\tt MESA} & {\tt MONSTAR} & {\tt YAP} & {\tt YREC} \\
\hline
\smallskip
\smallskip
1.0 &  7.0 &  72.12 &  72.07 &  72.58 &  72.41 &  72.64 &  72.69 &  73.23 &  76.64 &  73.14\\
1.0 & 12.0 &  58.36 &  58.00 &  58.39 &  58.44 &  58.47 &  58.61 &  59.02 &  60.86 &  58.97\\
1.5 &  7.0 &  69.90 &  69.86 &  70.34 &  70.20 &  70.48 &  70.75 &  71.13 &  74.48 &  71.05\\
1.5 & 12.0 &  57.29 &  56.94 &  57.43 &  57.35 &  57.30 &  57.48 &  57.87 &  59.47 &  57.90\\
2.0 & 10.0 &  78.72 &  79.02 &  78.36 &  77.18 &  78.14 &  76.99 &  79.57 &  82.26 &  79.02\\
2.5 & 10.0 & 123.62 & 124.10 & 121.54 & 121.50 & 122.32 & 121.83 & 124.91 & 125.52 & 123.33\\
\hline
\end{tabular}
\end{table*}

\begin{figure}[ht]
\includegraphics[angle=0,scale=0.5]{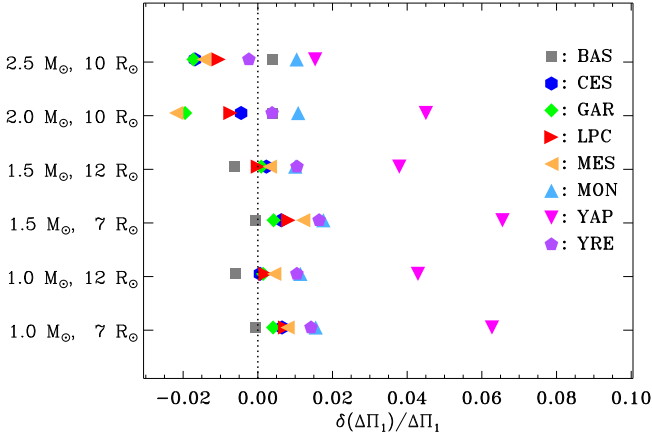}
\caption{
Relative differences for solar-calibrated models
in the asymptotic period spacing $\Delta \Pi_1$,
compared with the {\tt ASTEC} results, in the sense (model) - ({\tt ASTEC});
the different codes are identified by the symbol shape and colour 
(cf.\ caption to Fig.~\ref{fig:diffdr0_solar}).
}
\label{fig:diffdpisolar}
\end{figure}

The asymptotic dipolar g-mode period spacing $\Delta \Pi_1 = \Pi_0/\sqrt{2}$
(cf. Eqs.~\ref{eq:gasymp} and \ref{eq:dpi})
are provided in Table~\ref{tab:perspacsolar}
for solar-calibrated models, while the variations relative to the
{\tt ASTEC} models are shown in Fig.~\ref{fig:diffdpisolar}.
Here relative differences of up to 2 -- 4\,\% are found, corresponding to
differences in $\Delta \Pi_1$ of several seconds, greatly exceeding
the observational uncertainty of around 0.1 s (see Section~\ref{sec:obsprop}).
This reflects the sensitivity
of the buoyancy frequency to the details of the core structure of the star,
including the composition profile;
we consider one example in some detail in Appendix~\ref{sec:astccore}.
The differences between the models illustrated in Fig.~\ref{fig:diffdpisolar}
generally arise from qualitatively similar, although generally smaller,
model differences.

\begin{figure}[ht]
\includegraphics[angle=0,scale=0.5]{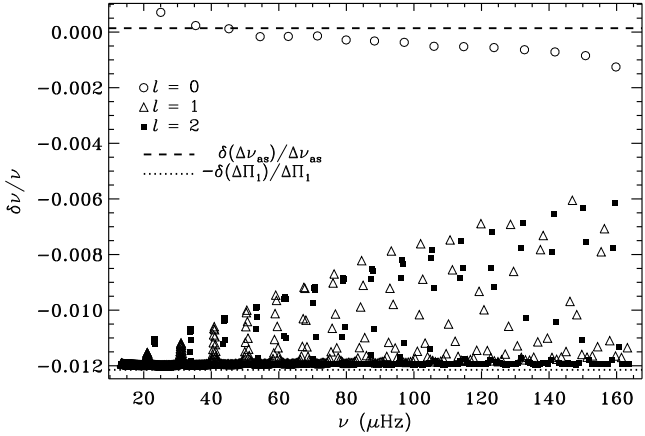}
\caption{
Relative differences in computed frequencies for the
{\tt MESA} solar-calibrated $1.5 \msun, 7 \rsun$ model,
compared with the {\tt ASTEC} results, in the sense
({\tt MESA}) - ({\tt ASTEC}),
for $l = 0$ (open circles), $l = 1$ (open triangles) and $l = 2$
(filled squares).
The differences are evaluated at fixed radial order.
}
\label{fig:dfmesm15r7}
\end{figure}

\begin{figure}[ht]
\includegraphics[angle=0,scale=0.5]{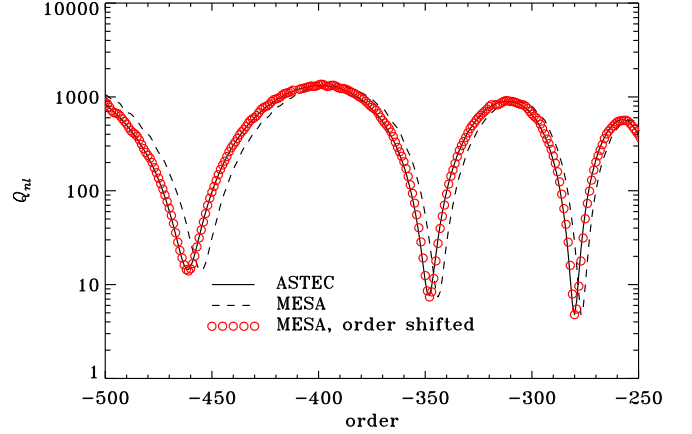}
\caption{
Scaled mode inertias for dipolar modes (with $l = 1$)
in the solar-calibrated {\tt ASTEC} (solid) and {\tt MESA} (dashed) $1.5 \msun, 7 \rsun$ models, against mode order.
The red circles show the {\tt MESA} results, but shifted in order (see text).
}
\label{fig:qcomp}
\end{figure}



As an example of the differences in individual frequencies,
Fig.~\ref{fig:dfmesm15r7} compares frequencies of given
radial order in the {\tt MESA} model for $1.5 \msun, 7 \rsun$ with {\tt ASTEC}.
The difference in the asymptotic frequency spacing is shown as a dashed line,
and the dotted line shows the difference in the asymptotic period spacing,
with inverted sign to convert relative period differences
to frequency differences.
In this case, the purely acoustic radial modes generally agree well between
the two models, as does the asymptotic frequency spacing
(see also Fig.~\ref{fig:diffdnusolar}).
However, we note the increasing magnitude of the differences at the
highest frequency that reflects issues with the modelling of the 
atmosphere in the MESA model (see Appendix~\ref{sec:mesatm}).
The g-dominated nonradial modes have differences very close to the
asymptotic value, while the more acoustically dominated modes have intermediate 
differences.
Indeed, one would naively expect that the p-dominated nonradial
modes would have frequency differences similar to the radial modes;
instead, they are substantially higher in absolute value,
with a clear nearly linear envelope for the most p-dominated cases.

The origin of this behaviour lies in the formally reasonable choice
of comparing the modes at fixed radial order, regardless of the physical
nature of the modes.
In fact, modes with same order may have rather different physical nature.
To analyse this, we consider the
rescaled inertia $Q_{nl} = E_{nl}/\bar E_0(\nu_{nl})$ where $E$ was defined in
Eq.~(\ref{eq:inertia}) and $\bar E_0(\nu_{nl})$ is the radial-mode inertia
interpolated to the frequency of the given mode.
Figure~\ref{fig:qcomp} shows $Q_{nl}$ for $l = 1$ against mode order
for the two models.
Here the p-dominated modes correspond to the dips in the curves,
resulting from acoustic resonances.
The resonances are largely fixed at the same frequency by the
very similar acoustic behaviour of the {\tt MESA} and {\tt ASTEC}
models,
reflected in the close agreement in the radial-mode frequencies;
however, it is obvious that they are shifted in mode order,
as a result of the difference between the models
in the period spacing and hence the relation between order and frequency.
In other words, although the two models agree on the shape 
of the $Q_{nl}$ curve 
(including the location in frequency of the resonant dips),
the mixed modes of the two models do not sample that curve 
at the same mode orders.
As a result, a comparison at fixed order is between physically different
modes, with a different weight to the p- and g-mode behaviour,
in the vicinity of the acoustic resonances.

\begin{figure}[ht]
\includegraphics[angle=0,scale=0.5]{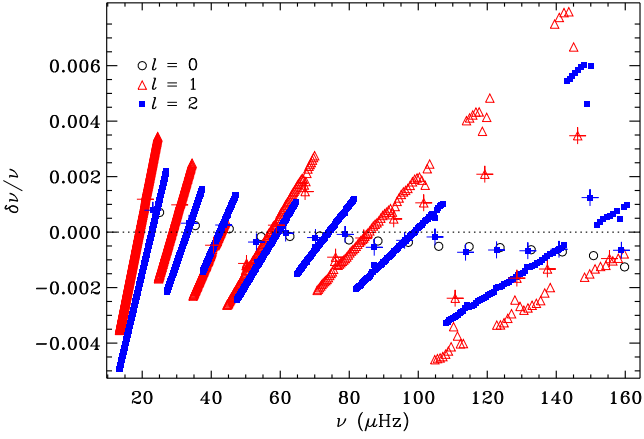}
\caption{
Relative differences in computed frequencies for the solar-calibrated
{\tt MESA} $1.5 \msun, 7 \rsun$ model with suitable shifts in mode order
(see text), compared with the {\tt ASTEC} results, in the sense
({\tt MESA}) - ({\tt ASTEC}),
for $l = 0$ (black open circles), $l = 1$ (red open triangles) and $l = 2$
(blue filled squares).
The larger pluses mark the p-dominated modes.
The horizontal dotted line indicates zero difference.
}
\label{fig:dfmesm15r7shft}
\end{figure}

From the point of view of comparing models and observations, the (formal)
order is a somewhat inconvenient quantity since it is difficult to
derive it directly from the observations,
except with data of exceptionally high quality.
Here the p-dominated modes are the natural starting points, 
anchoring the mode orders in the vicinity of an acoustic resonance.
In the comparison of the model frequencies, this corresponds to
shifting the mode orders of, say, the {\tt MESA}
model to obtain a new order $n'$ such as to make the acoustic resonances occur
at the same values of the order.
As indicated by Fig.~\ref{fig:qcomp}, the required shift decreases with
increasing order.
Thus, in the complete set $\{n'\}$ of shifted orders there may be gaps
or overlapping modes, but these can be arranged to occur near the maxima
in $Q_{nl}$ where the modes are unlikely to be observed.
As shown by the red circles in Fig.~\ref{fig:qcomp}, with such a shift the
behaviour of $Q_{nl}$ as a function of mode order
is nearly indistinguishable between the models. 

The effect of using the shifted orders in the frequency comparison
at fixed order is illustrated in Fig.~\ref{fig:dfmesm15r7shft}.
Now the p-dominated modes, marked by larger pluses superposed on the symbols,
do indeed have small frequency differences.
This is particularly clear for the $l = 2$ modes, where the coupling 
between the acoustic and gravity-wave regions is weaker and the p-dominated
modes therefore have a cleaner acoustic nature. 

To understand the behaviour shown in Fig.~\ref{fig:dfmesm15r7shft}, we consider
two models, Model 1 (the {\tt ASTEC} model) and Model 2 (the {\tt MESA} model),
with period spacings $\Delta \Pi_l$ and $\Delta \Pi_l'$.
For simplicity, we assume that the g-mode phase shift $\epsilon_{\rm g}$
(cf.\ Eq.~\ref{eq:gasymp}) is the same for the two models.
We identify an acoustic resonance in Model 1,
corresponding to the order $n_0$, and consider modes of
order $n$ in Model 1 in the vicinity of $n_0$.
To identify modes in Model 2 similarly close to the acoustic resonance,
we choose a shift $k$ in order such that
$\Delta \Pi_l(n_0 + \epsilon_{\rm g})
\simeq \Delta \Pi_l'(n_0 + k + \epsilon_{\rm g})$, or
\begin{equation}
k \simeq - {\delta \Delta \Pi_l \over \Delta \Pi_l'} n_0 \; ,
\label{eq:shift}
\end{equation}
where $\delta \Delta \Pi_l = \Delta \Pi_l' - \Delta \Pi_l$,
and compare modes with shifted mode order $n' = n+k$ in Model 2 
with modes of order $n$ in Model 1.
From Eq.~(\ref{eq:shift}) it follows that
the relative difference between the frequencies 
$\nu_{n'l}'$ and $\nu_{nl}$ of modes $(n',l)$ and $(n,l)$ in Models 2 and~1 is
\begin{eqnarray}
\label{eq:frqdif}
{\nu_{n'l}' - \nu_{nl} \over \nu_{nl}} &=&
- {\Pi_{n'l}' - \Pi_{nl} \over \Pi_{nl}} 
\simeq - \delta \Delta \Pi_l { n - n_0 \over \Pi_{nl}} \\
&\simeq& - {\delta \Delta \Pi_l \over \Delta \Pi_l}
{\Pi_{nl} - \Pi_{n_0 l} \over  \Pi_{nl}} 
\simeq {\delta \Delta \Pi_l \over \Delta \Pi_l}
\left({\nu_{nl} \over \nu_{n_0 l}} - 1\right) \nonumber \; ,
\end{eqnarray}
where we neglected the difference between $\Delta \Pi_l'$ 
and $\Delta \Pi_l$ in the denominator.
Equation~(\ref{eq:dpi}) shows that
$\delta \Delta \Pi_l /\Delta \Pi_l$ is independent of $l$.
Thus, according to Eq.~(\ref{eq:frqdif}) the
frequency differences including the shift in mode order
are linear functions of frequency with a slope
depending on $\delta \Delta \Pi_l /\Delta \Pi_l$ and $\nu_{n_0 l}$
but not on the degree, as is indeed found in Fig.~\ref{fig:dfmesm15r7shft}.



The detailed frequency differences for other models or evolution codes
are qualitatively similar, although reflecting the differences in global
asteroseismic properties, in particular $\Delta \nu_{\rm as}$ and
$\Delta \Pi_1$, as illustrated in Figs.~\ref{fig:diffdnusolar} 
and \ref{fig:diffdpisolar}.
However, we note that these models to some extent reflect modifications
to the modelling codes resulting from the analysis of earlier models.
Earlier models, showing substantially larger deviations, provide 
interesting insight into the relation between the model structure and
the resulting frequencies.
We discuss examples of this in Appendices
\ref{sec:asteos} and \ref{sec:astccore}.

\begin{figure}[ht]
\includegraphics[angle=0,scale=0.5]{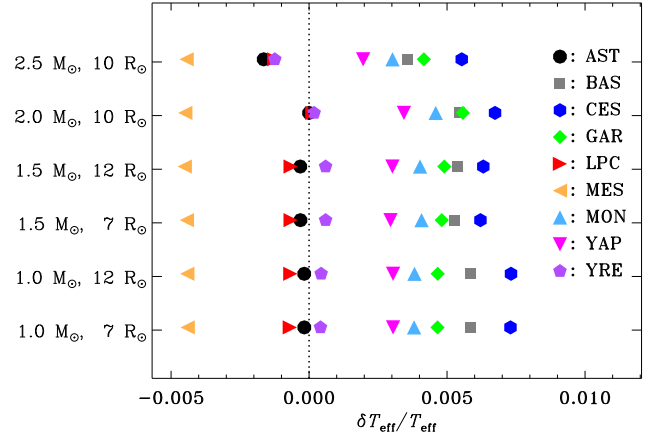}
\caption{
Relative differences in effective temperature $T_{\rm eff}$ between the
RGB- and solar-calibrated models, in the sense
(RGB-calibrated) -- (solar-calibrated).
The different codes are identified by the symbol shape and colour 
(cf.\ caption to Fig.~\ref{fig:diffdr0_solar}).
}
\label{fig:diffteff}
\end{figure}

\subsection{The RGB-calibrated models}
\label{sec:rgbdiscuss}

In Paper~I, we showed that the effective temperature on the red-giant branch
varied within a range of around 50\,K between the solar-calibrated models.
This variation is summarised in Fig.~\ref{fig:diffteff}, which shows
the differences between the effective temperatures in 
the RGB- and solar-calibrated models.
(The differences are small for the {\tt ASTEC} models, earlier versions of
which were used to set the target values for the RGB calibration.)
As discussed in Section~\ref{sec:acoustic}, $T_{\rm eff}$ directly
enters $\nu_{\rm max}$ 
and hence shows a much better agreement in the RGB- than in the solar-calibrated
case (cf.\ Fig.~\ref{fig:diffnumax});
in particular, the differences in the latter case closely reflect the
differences in $T_{\rm eff}$, shown in Fig.~\ref{fig:diffteff}.

Other results for the RGB-calibrated modes are provided as tables and figures
in Appendix \ref{sec:rgbcal}.
A comparison between the various asteroseismic quantities between the
RGB- and solar-calibrated cases (not illustrated) shows differences that
are to some extent, but not completely, 
related to the differences in $T_{\rm eff}$
and are somewhat smaller than the differences between the different
model calculations.
Consequently, the variations between the codes in the RGB-calibrated case
are qualitatively very similar to the variations discussed in the
previous sections.

\section{Discussion}

The goal of the present project is to provide a secure basis for the
analysis of observed frequencies of red-giant stars by identifying and
eliminating errors and other uncertainties in the computation of stellar 
models and their frequencies.
Paper~I considered differences between different stellar evolution codes
in the basic properties of stellar models, computed with tightly 
constrained parameters and physics.
Here we address the corresponding properties of the oscillations of
these models.

The errors in the computed frequencies include intrinsic errors
in the frequency calculation, for a given stellar model.
These are relatively easy to control, at least for the cases considered in
the present investigation.
The example illustrated in Fig.~\ref{fig:freqerr} indicates that the 
intrinsic numerical errors are well below the requirements imposed by current 
observations in the relevant frequency range 
for the oscillation code used here.
Even so, there is a definite need for comparisons, planned in a future
publication, between the results of independent oscillation calculations
to detect possible systematic errors in the implementation
of the oscillation equations in the codes.

A more important contribution to the errors in the computed frequencies
is likely the error in the implementation and solution of the equations
of stellar structure and evolution.
The goal of the present investigation is to estimate these errors,
by comparing results of stellar evolution calculations with independent
codes using, as far as possible, the same physical assumptions
(see Paper~I).
In fact, the analysis in the present project has identified, and led to the
elimination of, a number of issues
that have in some cases been present for many years
in the codes that have taken part.
Even so, following these corrections
the results in Section~\ref{sec:res} show that the model differences,
reflecting potential errors in the modelling, in many cases do not yet
match the constraints of the observational uncertainties.
The estimates of $\nu_{\rm max}$ (cf.\ Table~\ref{tab:numaxsolar}
and Fig.~\ref{fig:diffnumax}) agree substantially better than the 
relatively large observational uncertainty in this quantity.
The differences in the large frequency separation $\Delta \nu_{\rm fit}$
obtained from fits to the radial-mode frequencies
(cf.\ Table~\ref{tab:dnufrqsolar} and Fig.~\ref{fig:diffdnufrqsolar}) 
are close to matching, or somewhat exceed, the accuracy of the observations.
Finally, the root-mean-square differences of radial-mode frequencies 
(Fig.~\ref{fig:diffdr0_solar}) are substantially bigger than the
observational uncertainties in individual frequencies.

For the use of scaling relations based on acoustic modes, the correction
factor $f_{\rm \Delta \nu}$ (cf.\ Eq.~\ref{eq:dnucalscale})
is particularly important.
The spread in $f _{\Delta \nu}$ of $\pm 0.2$\,\% between the different
evolution codes (see Eq.~\ref{fig:diffdnusclsolar})
translate into variations of around 0.4 and 0.8\,\%
in determinations of radius and mass from asteroseismic scaling relations,
which are hardly insignificant.

Analyses of the properties of mixed modes provide detailed diagnostics
of the deep interior of the star, owing to the sensitivity of the details
of the acoustic resonances and the g-dominated modes.
The analysis is often carried out in terms of fits of the frequencies
to the asymptotic expression \citep[e.g.][]{Mosser2018},
resulting in estimates of the g-mode period spacing $\Delta \Pi_1$,
the quantity $q$ characterising the coupling between the g- and p-mode
cavities and the gravity offset $\epsilon_{\rm g}$.
Here we represented the effects of the model differences in terms
of the asymptotic period spacing (cf.\ Eq.~\ref{eq:dpi}).
A more detailed analysis in terms of a fit to the computed frequencies would
have been interesting but is beyond the scope of this paper.
However, a sample check for a single model case showed
that the asymptotic period spacings are fully representative of the results
based on period spacings obtained from such a fit.
We note, on the other hand, that very interesting analyses of the information 
about stellar structure provided by $q$ and $\epsilon_{\rm g}$ were provided by
\citet{Takata2016, Pincon2019}.

The sensitivity of the computed asymptotic dipolar period spacings
(cf.\ Table~\ref{tab:perspacsolar} and Fig.~\ref{fig:diffdpisolar})
to the detailed structure of the deep interior of the models
is reflected in a substantial spread between the models, 
far bigger than the observational errors.
In most cases, the relative differences are between $\pm 2$\,\%, 
with the {\tt YaPSI} models showing somewhat bigger deviations.
As a rough estimate we note from Table~\ref{tab:perspacsolar} that
over the range $1 - 1.5 \msun$ in stellar mass,
keeping the radius fixed at $7 \rsun$, a change of one per~cent in
$\Delta \Pi_1$ corresponds on average to a change of more 
than $0.1 \msun$ in mass
or a change in the inferred age of more than 30\,\%.
Even though a model fit based solely on $\Delta \Pi_1$ is probably
unrealistic, this estimate provides some indication of the effects of the 
uncertainties in stellar modelling on the asteroseismic inferences.

These differences in oscillation properties must reflect differences
in the model structure, discussed in detail in Paper~I, 
which arise despite the attempt to compute the models under identical
assumptions;
however, the connection is in most cases not immediately obvious.
We analysed two examples in some detail.
Appendix \ref{sec:asteos} 
considers differences in the acoustic-mode frequencies 
in the original {\tt GARSTEC} models (cf.\ Fig.~\ref{fig:dfgarm15r7}),
which were found to be caused by differences in
the implementation of the OPAL equation of state, illustrated in 
Fig.~\ref{fig:diffcsq}.
Appendix~\ref{sec:astccore} analyses the fairly substantial differences 
found in the asymptotic $\Delta \Pi_1$ and the g-m mode frequencies
for the original {\tt LPCODE} model with $2.5 \msun$, $10 \rsun$.
As discussed in detail in the appendix,
this is related to differences in the hydrogen profile arising from
a smaller main-sequence convective core in the {\tt LPCODE} model,
caused by inadequacies in the opacities.
This deficiency has been corrected in the {\tt LPCODE} results shown 
in Section~\ref{sec:res},
as perhaps the most dramatic of the many corrections to the modelling 
resulting from this challenge.
It should be noted that the oscillation calculations act as a
strong `magnifying glass' on irregularities in the model structure,
further motivating such improvements to the modelling;
an example is discussed in Section~\ref{sec:diporder}.

In the analysis of the results, we chose to emphasise the case
of models where the mixing-length parameter was chosen based on the
calibration of a $1 \msun, 1 \rsun$ model (the so-called solar-calibrated case).
This procedure matches the common practice of using 
such a calibration in general calculations of stellar models,
including those that are used for asteroseismic fitting.
From a physical point of view one might argue that the RGB-calibration,
based on fixing the effective temperature on the red-giant branch,
is more interesting since by doing this (at the assumed fixed radii)
one also fixes the luminosity and hence important aspects of the internal
structure of the stars.
In fact, the results for the two different calibrations are quite similar,
and hence the choice does not affect the overall conclusions of this study.

\section{Conclusions}

The huge amount of high-accuracy oscillation data resulting from the 
{\it Kepler} mission, 
which is currently being augmented by the ongoing TESS mission,
provides an opportunity to investigate stellar properties in considerable
detail, 
thereby helping to improve our understanding of stellar structure and evolution.
The use of observed oscillation frequencies as diagnostics of stellar
global and internal properties in most cases relies on the comparison with
frequencies of stellar models.
For this to be meaningful and hence ideally to
utilise fully the accuracy provided by the observed frequencies,
the numerical errors in the computed
frequencies should be constrained, in principle to be well below
the observational uncertainties. 
With data of the quality obtained from the {\it Kepler} mission,
this is an ambitious goal.

The analyses presented in this paper and Paper~I
represent a significant step towards a coordinated and coherent 
modelling of stars and their oscillation frequencies.
Compared with other common uses of stellar modelling,
such as diagnostics based on observed properties of colour-magnitude diagrams
or isochrone fitting,
the results obtained here already demonstrate a reasonable convergence towards
consistent stellar models for given physics, 
with differences at the level of a few tenths of and up to a few per cent.
However, continuing efforts will be required to investigate the remaining
differences in the individual cases, starting with the differences in
the results of the evolution modelling,
and the possible required further improvements to the codes.
We hope that by presenting the results in some detail in the present paper
and as an on-line resource, they can also serve as useful references in
comparisons with other codes that have not been involved in the
present project or in the development of techniques for the analysis
of observational data.

The sensitivity of the frequencies to even quite small details in the models 
demonstrates the potential of the oscillation data for probing 
subtle features of the stellar interiors.
This will be further explored in a future publication, where the
modellers will consider individually selected physical properties
of the models, moving closer to the realistic modelling to be used
in fits of the observed data.
Based on these efforts, we expect to be in a better position to interpret
the results of such fits in terms of the physics of stellar interiors, 
which, after all, is an important goal of asteroseismic investigations.
Also, we hope that the investigations will help in improving 
the understanding of and reducing
the systematic errors in the resulting global stellar
properties inferred from asteroseismology, in particular the age.
This is an important part of the analysis of existing data and, 
in particular, the preparation for the upcoming ESA PLATO mission
\citep[e.g.][]{Rauer2014}, where asteroseismic stellar characterisation is a
key part of the data analysis.

\begin{acknowledgements}
We are grateful to Ian Roxburgh for pointing out to us the problems
with the atmospheric structure in the MESA models.
The referee is thanked for constructive comments which have substantially
improved the presentation.
Funding for the Stellar Astrophysics Centre is provided by
The Danish National Research Foundation (Grant agreement No. DNRF106).
The research was supported by the ASTERISK project
(ASTERoseismic Investigations with SONG and {\it Kepler})
funded by the European Research Council (Grant agreement No. 267864).
This research was supported in part by the National Science Foundation
under Grant No. NSF PHY-1748958.
VSA acknowledges support from VILLUM FONDEN (research grant 10118) 
and the Independent Research Fund Denmark (Research grant 7027-00096B).
DS is the recipient of an Australian Research Council Future
Fellowship (project number FT1400147).
SC acknowledges support from Premiale INAF MITiC, from INAF 
`Progetto mainstream' (PI: S. Cassisi),
and grant AYA2013-42781P from the Ministry of Economy and Competitiveness
of Spain.
AMS is partially supported by grants ESP2017-82674-R (Spanish Government)
and 2017-SGR-1131 (Generalitat de Catalunya).
TC acknowledges support from the European Research Council AdG No
320478-TOFU and the STFC Consolidated Grant ST/R000395/1.
SH received funding for this research from the European Research Council under the European Community's Seventh Framework Programme (FP7/2007-2013) / ERC grant agreement no 338251 (StellarAges).
AM acknowledges the support of the Government of India,
Department of Atomic Energy, under Project No. 12-R\&D-TFR-6.04-0600.
\end{acknowledgements}
\bibliographystyle{aa} 

%
%
\begin{appendix}
\section{Frequency calculations}\label{sec:freqcal}

\subsection{Computational procedures}
\label{sec:compproc}

The models were provided by the participants in the so-called
{\tt fgong} format, which includes a substantial number of model variables
at all meshpoints in the evolution computation, together with global
parameters.
The model is transferred to the {\tt amdl} format required for the calculation
of adiabatic frequencies.
Subsequently, the model is moved to a new mesh optimised for the 
frequency calculation, which is then carried out by the {\tt ADIPLS} code
\citep[cf.][]{Christ2008a}, with output both in binary form and
in the form of an ASCII {\tt fobs} file.
In the following, we describe each of these steps in a little more detail.

Models computed with general stellar evolution codes sometimes contain
features of little importance to general stellar evolution but harmful 
for oscillation calculations.
Such problems in particular concern the Ledoux discriminant
\begin{equation}
A = {1 \over \Gamma_1} {\dd \ln p \over \dd \ln r} 
- {\dd \ln \rho \over \dd \ln r} \; ,
\label{eq:ledoux}
\end{equation}
related to the buoyancy frequency by $N^2 = g A/r$,
which is highly sensitive to irregularities in the composition profile.
Particularly harmful are negative spikes in $A$ in the stellar core,
where $g$ is large, which leads to (unrealistically) strong convective
instability.
In the transfer to the {\tt amdl} format, such spikes are simply replaced
by interpolation from neighbouring points, setting $A = 0$ if the result is
negative.
We note that such resetting of $A$ without corresponding changes to other
variables formally leads to inconsistency in the model, 
a point that deserves further attention.
The models are tested for double points, with identical $r$ at the
accuracy of the model format, and such points are removed,
except if they are associated with discontinuities in the model structure
(see below).
Finally, the oscillation calculation requires second derivatives of $p$
and $\rho$ at $r = 0$;
if these are not available in the original model they are estimated from the
behaviour of these quantities near the central meshpoint.

In the relevant frequency range in red giants, the number of radial nodes
in the g-mode region may exceed 1000, requiring a very dense radial
mesh to resolve the eigenfunctions.
This is, in general, not satisfied by the mesh in the evolution calculation,
requiring for the model to be transferred to a new mesh with a higher number
of points and an appropriate distribution.
Guidance for the mesh distribution follows from the asymptotic behaviour
of the modes \citep[see also][]{Hekker2017}.
In the g-mode region, where the modes behave as internal gravity waves,
the eigenfunction varies approximately as
\begin{equation}
\CA_{\rm g} (r) \sin\left( {L \over \omega} \upsilon \right) \; ,
\label{eq:gmode}
\end{equation}
where $L = \sqrt{l(l+1)}$ and
\begin{equation}
\upsilon = \int_0^r N {\dd r \over r}
\end{equation}
is the buoyancy radius.
The predominantly acoustic behaviour in the p-mode region has the form
\begin{equation}
\CA_{\rm p} (r) \sin ( \omega \tau ) \; ,
\label{eq:pmode}
\end{equation}
where 
\begin{equation}
\tau = \int_r^R {\dd r \over c} 
\end{equation}
is the acoustic depth.
In Eqs.~(\ref{eq:gmode}) and (\ref{eq:pmode})
$\CA_{\rm g}$ and $\CA_{\rm p}$ are slowly varying amplitude functions.
Thus, a reasonable distribution of the mesh involves approximately 
uniform spacing in $\upsilon$ and $\tau$ in the g- and p-mode regions,
respectively, with a suitable distribution in the intermediate region.
The appropriate balance between the relative number of points
assigned to the g- and p-mode regions
can be determined from the asymptotic analysis, given the frequency range
to be considered.

\begin{figure}[ht]
\includegraphics[angle=0,scale=0.5]{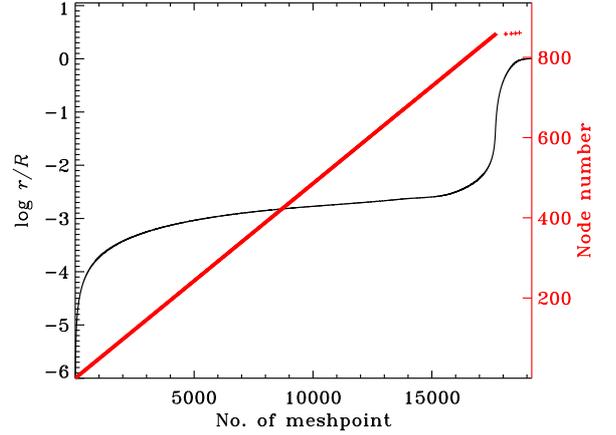}
\caption{
Properties of the mesh used for the oscillation calculation
in the $1.0 \msun, 12 \rsun$ case.
The solid black line shows the logarithm of the fractional radius,
against the mesh-point number (left ordinate scale).
The red crosses show the nodes in the horizontal-displacement eigenfunction
in a dipolar mode with frequency $20.0 \muHz$ 
(right ordinate scale).
}
\label{fig:mesh}
\end{figure}

In the present calculations, a mesh with 19\,200 points was used. 
The properties of the mesh are illustrated in Fig.~\ref{fig:mesh},
which shows the fractional radius against the mesh-point number.
It is evident that by far the majority of the points are in the core,
within $3 \times 10^{-3} R$, to match the g-mode-like behaviour in
this region.
Also shown are the locations of the nodes in the horizontal-displacement
eigenfunction. 
In the g-mode region these are almost uniformly spaced, with approximately
20 meshpoints between adjacent nodes.
The comparatively few nodes in the p-mode region have a wider spacing,
the mesh satisfying also the requirement of adequately resolving the
variation in the overall amplitude of the eigenfunctions.

Adiabatic oscillations satisfy a fourth-order system of equations.
Boundary conditions at $r = 0$ are defined by regularity conditions.
At the outermost meshpoint, one boundary condition is obtained from the
continuity of the perturbation to the gravitational potential and its gradient
and a second from requiring that the solution transits continuously to
the analytical solution of the adiabatic oscillation equations
in an assumed isothermal atmosphere 
continuously matched to the model at the outermost point.
The oscillation equations were solved using a fourth-order numerical scheme
\citep{Cash1980}.
The eigenfrequencies were obtained from the condition of continuous
matching of solutions integrated from the surface and the centre,
at a suitable point in the core.
This was achieved through a careful scan in frequency, reflecting the
asymptotic distribution of frequencies, to ensure that no modes were missed.
A test of the completeness was carried out on the basis of the
mode orders, determined as discussed in Section \ref{sec:diporder}.

\begin{figure}[ht]
\includegraphics[angle=0,scale=0.5]{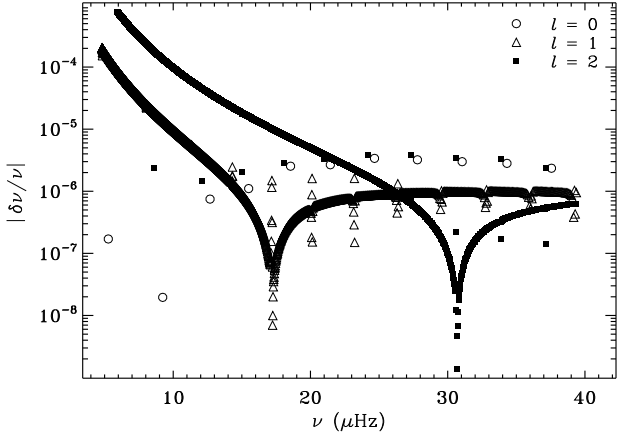}
\caption{
Absolute value of relative frequency differences,
at fixed radial order, between
computations with 38\,400 and 19\,200 points 
for the $1.0 \msun, 12 \rsun$ case.
}
\label{fig:freqerr}
\end{figure}

A special problem concerns discontinuities in composition and hence density,
which give rise to a delta-function behaviour of $A$ (cf. Eq.~\ref{eq:ledoux})
and, hence, in the buoyancy frequency. This occurs, for example, at the
edge of the dredge-up region caused by the convective envelope, given that
these models do not include diffusion and settling.
A proper treatment in the model of a discontinuity would be to include it as
a double point, at the same values of the continuous variables, 
but more typically it appears as a rapid variation in composition and
density between adjacent meshpoints.
A discontinuity in the model gives rise to discontinuities 
in the eigenfunctions, and these should ideally be dealt with by
solving the equations separately on the regions separated
by the discontinuities, applying jump conditions on the solution 
at these points.
In the present calculations, each density discontinuity was replaced 
in the code resetting the mesh by a thin region with
a very steep linear density gradient, fully resolved,
and the oscillation equations were solved across this region;
we ensured that the integral over the region of $A$,
represented as a box function, was consistent with the jump in density.
We confirmed that the relevant jump conditions on the eigenfunctions
are satisfied to adequate accuracy at these points.
One remaining issue is that the variation in composition is not adequately
resolved in some of the models included in the comparison.
In these cases, further resetting of the model (or, ideally, improvements to
the evolution codes) would be desirable,
and the treatment of such features in the model will also be a topic
in future development and comparisons of oscillation codes.

\begin{figure}[ht]
\includegraphics[angle=0,scale=0.5]{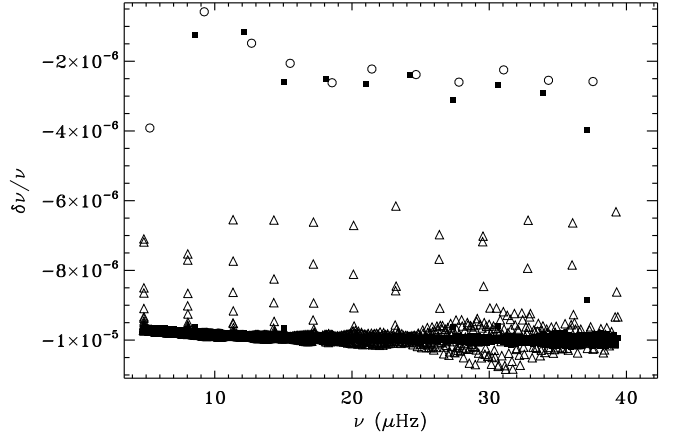}
\caption{
Relative frequency differences at fixed radial order
for $1 \msun, 12 \rsun$, between
an {\tt ASTEC} model using 8509 timesteps from the ZAMS
and the model used as reference with 4250 timesteps.
For symbol types, see Fig.~\ref{fig:freqerr}.
}
\label{fig:freqerrtime}
\end{figure}

\subsection{Numerical precision}
\label{sec:numprec}
Given the rapid variation in the eigenfunctions, the numerical accuracy 
is a concern, even given the precautions discussed above.
As a test of the accuracy, we computed frequencies for all {\tt ASTEC} models,
doubling the number of meshpoints; 
the differences between the original and refined computations then give
a measure of the numerical error in the former.
Figure~\ref{fig:freqerr} shows the results in the worst case, the most
evolved $1 \msun$ model.
Except for g-dominated modes of degree $l = 2$ at relatively low frequency,
the relative errors are generally below $10^{-4}$;
for the radial and p-dominated nonradial modes the errors are below $10^{-5}$,
with the effects of the p-m nature being particularly visible for $l = 1$.
We also note that, given that the models compared are very similar,
these numerical errors largely cancel in comparisons between 
models computed with different codes.
Thus, the results obtained in the main text are unaffected by numerical errors.
Even so, a comparison between different oscillation codes is obviously of
interest and is planned for a future publication.

A perhaps more serious issue is the numerical accuracy of the evolution
calculation resulting in the {\tt ASTEC} models that have been used as 
reference in the present frequency comparisons.
The models were computed with a fixed number of 1200 meshpoints,
whose distribution changes in response to the changing structure as the 
models evolve.
We verified that doubling the number of mesh points
in the evolution calculation has a negligible effect on the results.
The same is true of the number of timesteps
in the model calculation, which is controlled by a parameter
determining the maximum allowed change between two successive timesteps
in a suitable number of model variables, throughout the stellar interior.
To illustrate this, Fig.~\ref{fig:freqerrtime} shows
relative frequency differences between a $1 \msun$ model requiring 8509
timesteps to reach the radius $12 \rsun$ and the reference case 
with half as many steps. 

\subsection{Large frequency separation from frequency fitting}
\label{sec:dnufit}

\begin{figure}[ht]
\includegraphics[angle=0,scale=0.5]{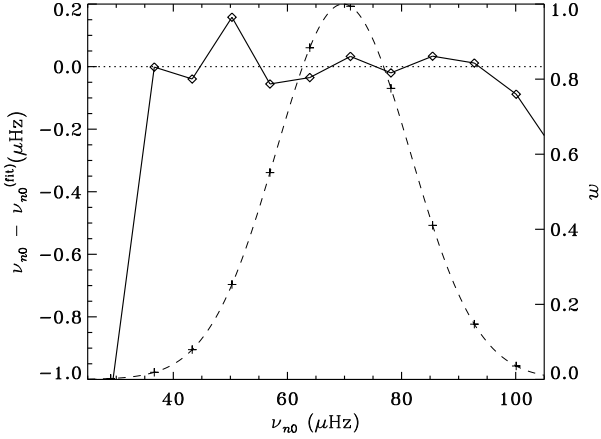}
\caption{
Determination of $\Delta \nu_{\rm fit}$ for the solar-calibrated
{\tt ASTEC} $1.0 \msun, 7 \rsun$ model.
The solid line shows the residual between the radial-mode frequencies
and the fitted function $\nu_{n0}^{\rm (fit)}$ (cf.\ Eq.~\ref{eq:fitfun}),
the diamonds indicating the location of the actual frequencies.
The dashed line shows the weight function $w$
(cf.\ Eq.~\ref{eq:fitweight}), using the right-hand ordinate.
}
\label{fig:fitdnu0}
\end{figure}

To determine $\Delta \nu_{\rm fit}$ we largely follow 
\citet{White2011} and carry out a weighted quadratic least-squares fit
of radial-mode frequencies $\nu_{n0}$ as functions of radial order $n$
by minimising
\begin{equation}
\Sigma_n w_n^2 (\nu_{n0}^{\rm (fit)} -\nu_{n0})^2 \; ,
\end{equation}
where
\begin{equation}
\nu_{n0}^{\rm (fit)} = \nu_0 
+ \Delta \nu_{\rm fit} [ (n - n_{\rm max})  + \alpha (n - n_{\rm max})^2] \; 
\label{eq:fitfun}
\end{equation}
\citep{Kjelds2005, Mosser2013}.
Here $\nu_0$ is a reference frequency and $n_{\rm max}$ is the
(generally non-integral) order corresponding to $\nu_{\rm max}$,
obtained by linear interpolation of $n$ as a function of $\nu_{n0}$.
Also,
\begin{equation}
w = \exp \left[ { - (\nu_{n0} - \nu_{\rm max})^2 \over 2 \sigma^2} \right] \, ,
\label{eq:fitweight}
\end{equation}
where
\begin{equation}
\sigma= \gamma { \nu_{\rm max} \over 2 \sqrt{2 \ln 2}} \; ,
\end{equation}
such that the full width at half maximum of $w$ is $\gamma \nu_{\rm max}$.
In the fits \citet{White2011} used $\gamma = 0.25$.
However, we instead followed \citet{Mosser2012} and 
evaluated $\gamma$ as 
\begin{equation}
\gamma = 0.66 (\nu_{\rm max}/1 \muHz)^{-0.12} \; ,
\end{equation}
based on a Gaussian approximation to the envelope of power;
this value of $\gamma$ changes 
from around 0.25 for the Sun to 0.45 for the $(1 \msun, 12 \rsun)$ models,
which have the lowest $\nu_{\rm max}$ (cf.\ Table~\ref{tab:numaxsolar}).
As an example, Fig.~\ref{fig:fitdnu0} shows the residuals from the fit
for the $1 \msun$, $7 \rsun$ {\tt ASTEC} model.

\begin{figure}[ht]
\includegraphics[angle=0,scale=0.5]{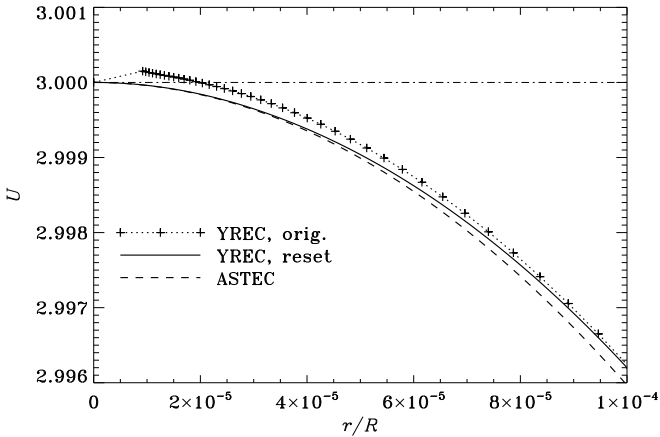}
\caption{
Behaviour of $U$ (cf. Eq.~\ref{eq:u}), in three models
with $M = 1 \msun$, $R = 7 \rsun$.
The dotted curve shows the original YREC model, the pluses marking
the mesh points in the evolution calculation.
The solid line shows the same model, after application of the correction
procedure discussed in the text.
For comparison, the dashed curve shows the corresponding {\tt ASTEC} model,
for which no correction had to be applied.
The thin dot-dashed line marks $U = 3$.
}
\label{fig:uprob}
\end{figure}

\subsection{Dipolar-mode order}
\label{sec:diporder}

Although the analysis of the frequency comparison 
in Section~\ref{sec:mixmodes} showed the limitations in using a formal
mode order not directly related to the physical nature of the mode,
a reliable formal determination of the mode order 
is an important feature of the frequency calculation. 
The order should be defined such that it is invariant for a given mode
as the star evolves. 
This, for example, allows reliable interpolation between frequencies
of modes at successive time steps in the model calculation.
Also, it has been applied to ensure that all modes have been found 
in the frequency ranges considered. 
Determination of a well-defined mode order for mixed modes requires that
the different characters of the eigenfunction 
in the g- and p-mode cavities is taken into account.
For modes of degree $l \ge 2$, this can be achieved by considering the
behaviour in a phase diagram defined by the vertical $\xi_r$ 
and horizontal $\xi_{\rm h}$ displacement amplitudes
\citep{Scufla1974, Osaki1975}.
The eigenfunction defines a curve in the $(\xi_r, \xi_{\rm h})$ diagram,
and a node in $\xi_r$ provides a positive (negative) contribution to
the mode order if the curve crosses the $\xi_r = 0$ axis in the 
counter-clockwise (clockwise) direction.

For centrally condensed stars, such as red giants, this procedure fails for
dipolar modes.
Following \citet{Takata2006}, we instead determined the order of such modes 
by means of a phase diagram based on 
\begin{equation}
{\cal Y}_1 = (3 - U) {\xi_r \over r} 
+ {1 \over g} \left( {\Phi' \over r} - {\dd \Phi' \over \dd r} \right) \; ,
\end{equation}
and
\begin{equation}
{\cal Y}_2 = (3 - U) {p' \over \rho g r} 
+ {1 \over g} \left( {\Phi' \over r} - {\dd \Phi' \over \dd r} \right) \; .
\end{equation}
Here $r$ is distance to the centre, 
$g$ is the local gravitational acceleration,
$\Phi'$ is the Eulerian perturbation to the gravitational potential and
$p'$ is the Eulerian pressure perturbation.
Also, 
\begin{equation}
U = {\dd \ln m \over \dd \ln r} = {4 \pi r^3 \rho \over m} \; ,
\label{eq:u}
\end{equation}
where $m$ is the mass internal to $r$.
As shown by Takata, and in general confirmed numerically, determining the mode
order based on zero crossings of ${\cal Y}_1$ and the direction of
rotation in the phase diagram provides a unique labelling of the modes.

The properties of ${\cal Y}_1$ and ${\cal Y}_2$ near $r = 0$ depend strongly
on the behaviour of $3 - U$.
Expanding $\rho$ to ${\cal O}(r^2)$ as
\begin{equation}
\rho = \rho_{\rm c}(1 - \varrho_2 r^2 + \ldots )  \; ,
\end{equation}
where $\rho_{\rm c}$ is the central density, we obtain
\begin{equation}
U = 3 \left( 1 - {2 \over 5} \varrho_2 r^2 + \ldots \right) \; .
\label{eq:uexp}
\end{equation}
Stability requires that $\rho$ decreases with increasing $r$,
and hence $\varrho_2 > 0$.
Thus, $3 - U$ tends smoothly to 0 for $r \rightarrow 0$ through positive values,
and the factor does not affect the topology of the first terms in ${\cal Y}_1$
and ${\cal Y}_2$.

Unfortunately, some of the models 
involved in the frequency comparison 
do not satisfy this behaviour of $U$ near the centre.
This is particularly serious when $U$, unphysically, exceeds 3, such
that $3 - U$ changes sign; this was the case for three codes.
An example is shown in Fig.~\ref{fig:uprob}; here $U$ exceeds 3 at the 
innermost points of the model resulting from the evolution code,
indicating an inconsistency in the way the inner boundary condition is applied.
If not corrected, this behaviour causes severe problems with the determination
of the order of dipolar modes.
For comparison the corresponding {\tt ASTEC} model is also shown;
here $U$ tends smoothly to 3 as $r \rightarrow 0$.
To secure a proper determination of the order, the problematic models
have been corrected in a manner that provides a reasonable behaviour
of $U$ near the centre.
Specifically, in models where $U$ exceeds 3 in the core the
outermost point $r_U$ where $U \ge 3$ was located. 
For $r \le r_U$, $U$ was reset to the result of the expansion, 
Eq.~(\ref{eq:uexp}), based on the expansion of $\rho$. 
On the interval $[r_U, 5 r_U]$ a gradual transition was made to
the original $U$, using a cubic polynomial determined such that $U$
and its first derivative are continuous.
The resulting corrected $U$ is also shown in Fig.~\ref{fig:uprob}.
This modification was applied in the code that transfers the original
model to a mesh suitable for the oscillation calculations
(see Section \ref{sec:compproc}).
To minimise the impact on the original models, the procedure was applied
only in cases where the uncorrected model was found to yield 
problematic dipolar mode orders.
These were identified as cases where one or more adjacent computed modes
did not correspond to mode orders differing by one.

The resetting of $U$ was carried out without any other readjustments of
the structure, thus raising legitimate concern about the internal consistency
of the resulting model. 
In fact, the computed frequencies for the reset and original models show
relative frequency differences of less than $10^{-6}$, and in almost all cases
less than $10^{-7}$, so that this has minimal consequences for the 
frequency comparisons carried out in the present paper.
Even so, it must clearly be a goal to revise the relevant modelling codes
to correct this problem at its root.
In general, the treatment of the innermost points in the model causes problems
in several cases, reflected in incorrect behaviour of $U$, 
although with no direct effect on the mode order;
in these cases no resetting of the model was carried out, and the
effects on the frequencies are likely insignificant, although again
revisions of the modelling codes are desirable.

\section{Results for the RGB-calibrated models}
\label{sec:rgbcal}

For completeness, we include a full set of results for the RGB-calibrated
models even though, as discussed in Section~\ref{sec:rgbdiscuss}, they are in most
cases very similar to those for the solar-calibrated models.

\begin{figure}[htpb]
\includegraphics[angle=0,scale=0.5]{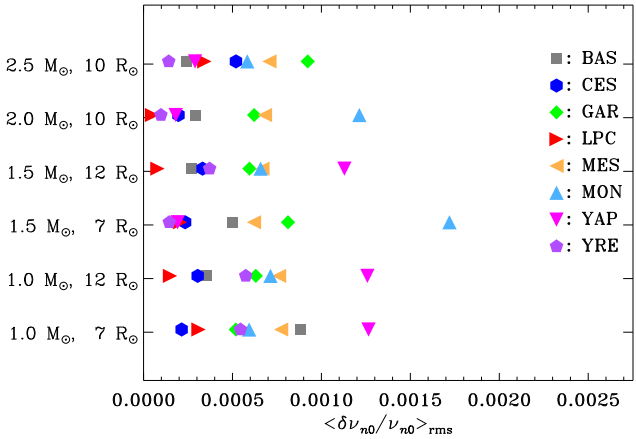}
\caption{
Root-mean-square relative differences for RGB-calibrated models
in radial-mode frequencies relative to
the {\tt ASTEC} results, in the sense (model) - ({\tt ASTEC});
the different codes are identified by the symbol shape and colour 
(cf.\ caption to Fig.~\ref{fig:diffdr0_solar}).
}
\label{fig:diffdr0_teff}
\end{figure}

\begin{table*}[htpb]
\caption{Large frequency separations $\Delta \nu_{\rm fit}$ in $\muHz$
obtained from fits to radial-mode frequencies as functions
of mode order (cf.\ Eq.~\ref{eq:pasymp} and Section \ref{sec:dnufit})
for RGB-calibrated models. }
\label{tab:dnufrqteff}
\centering
\begin{tabular}{r r c c c c c c c c c}
\hline\hline
$M/\msun$ & $R/\rsun$ & {\tt ASTEC} & {\tt BaSTI} & {\tt CESAM} & {\tt GARSTEC} & {\tt LPCODE} & {\tt MESA} & {\tt MONSTAR} & {\tt YAP} & {\tt YREC} \\
\hline
\smallskip
1.0 &  7.0 &  7.087 &  7.097 &  7.092 &  7.083 &  7.090 &  7.081 &  7.079 &  7.096 &  7.089\\
1.0 & 12.0 &  3.130 &  3.129 &  3.133 &  3.128 &  3.131 &  3.128 &  3.126 &  3.135 &  3.131\\
1.5 &  7.0 &  8.783 &  8.778 &  8.785 &  8.774 &  8.780 &  8.773 &  8.757 &  8.782 &  8.782\\
1.5 & 12.0 &  3.876 &  3.875 &  3.880 &  3.873 &  3.876 &  3.872 &  3.871 &  3.880 &  3.876\\
2.0 & 10.0 &  5.949 &  5.948 &  5.952 &  5.945 &  5.949 &  5.944 &  5.936 &  5.947 &  5.950\\
2.5 & 10.0 &  6.740 &  6.745 &  6.739 &  6.732 &  6.738 &  6.730 &  6.733 &  6.746 &  6.738\\
\hline
\end{tabular}
\end{table*}

\begin{figure}[htpb]
\includegraphics[angle=0,scale=0.5]{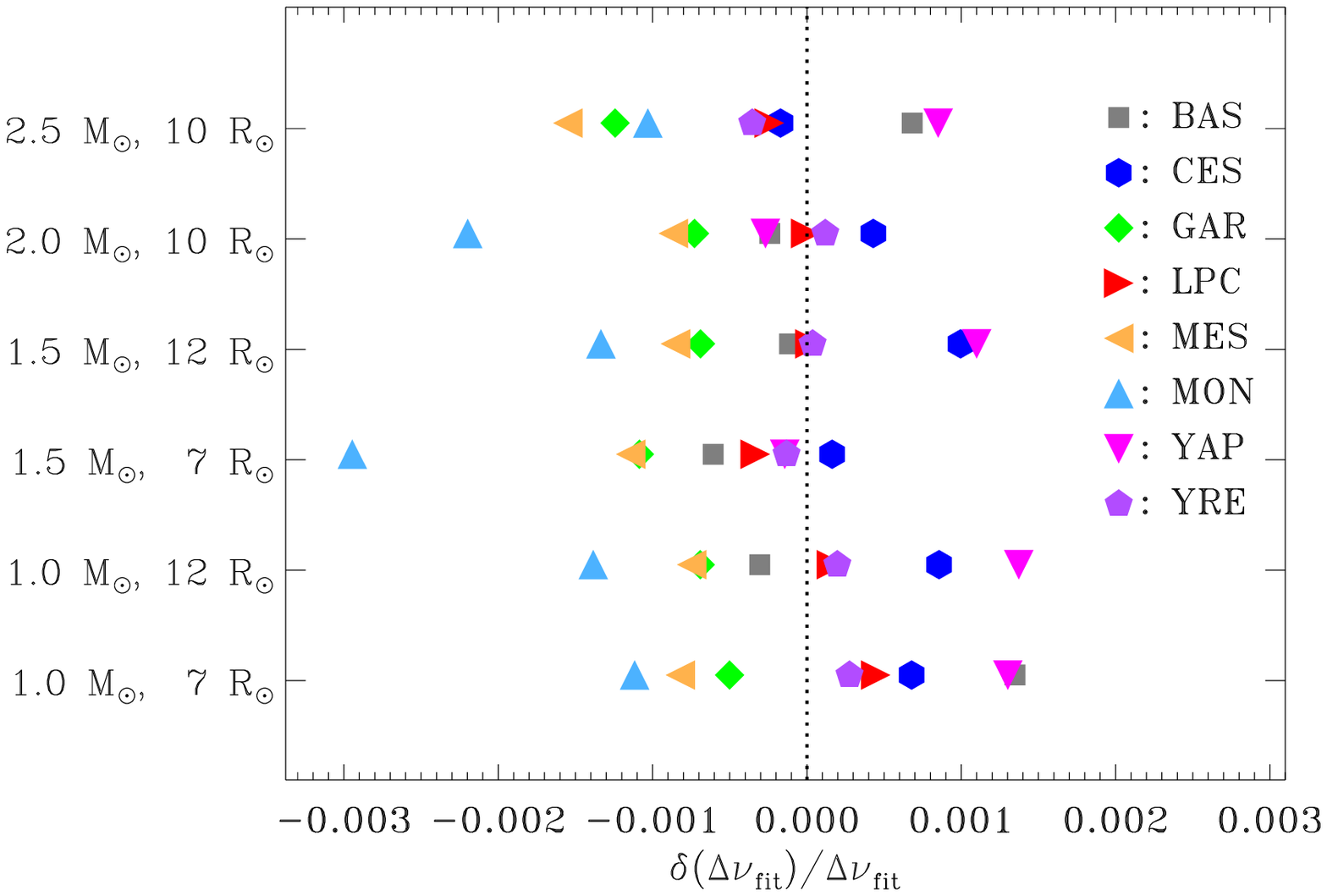}
\caption{
Relative differences for RGB-calibrated models
in the large frequency separations
$\Delta \nu_{\rm fit}$ obtained from fits to the radial-mode 
frequencies as functions of mode order (cf.\ Section \ref{sec:dnufit}),
compared with the {\tt ASTEC} results, in the sense (model) -- ({\tt ASTEC});
the different codes are identified by the symbol shape and colour 
(cf.\ caption to Fig.~\ref{fig:diffdr0_solar}).
\label{fig:diffdnufrqteff}
}
\end{figure}

\begin{table*}
\caption{Correction factors $f_{\Delta \nu}$ (cf.\ Eq.~\ref{eq:dnucalscale})
between the large frequency separation $\Delta \nu_{\rm fit}$ obtained from 
a fit to radial-mode frequencies and the value obtained
from homology scaling, for RGB-calibrated models.}
\label{tab:dnusclteff}
\centering
\begin{tabular}{r r c c c c c c c c c}
\hline\hline
$M/\msun$ & $R/\rsun$ & {\tt ASTEC} & {\tt BaSTI} & {\tt CESAM} & {\tt GARSTEC} & {\tt LPCODE} & {\tt MESA} & {\tt MONSTAR} & {\tt YAP} & {\tt YREC} \\
\hline
\smallskip
1.0 &  7.0 &  0.9656 &  0.9666 &  0.9656 &  0.9651 &  0.9660 &  0.9662 &  0.9650 &  0.9677 &  0.9657\\
1.0 & 12.0 &  0.9572 &  0.9567 &  0.9575 &  0.9566 &  0.9574 &  0.9579 &  0.9564 &  0.9594 &  0.9573\\
1.5 &  7.0 &  0.9771 &  0.9763 &  0.9766 &  0.9760 &  0.9768 &  0.9774 &  0.9747 &  0.9778 &  0.9768\\
1.5 & 12.0 &  0.9677 &  0.9674 &  0.9681 &  0.9671 &  0.9678 &  0.9683 &  0.9669 &  0.9697 &  0.9676\\
2.0 & 10.0 &  0.9786 &  0.9781 &  0.9784 &  0.9779 &  0.9786 &  0.9792 &  0.9769 &  0.9792 &  0.9785\\
2.5 & 10.0 &  0.9917 &  0.9921 &  0.9909 &  0.9905 &  0.9914 &  0.9916 &  0.9911 &  0.9934 &  0.9911\\
\hline
\end{tabular}
\end{table*}

\begin{figure}[htpb]
\includegraphics[angle=0,scale=0.5]{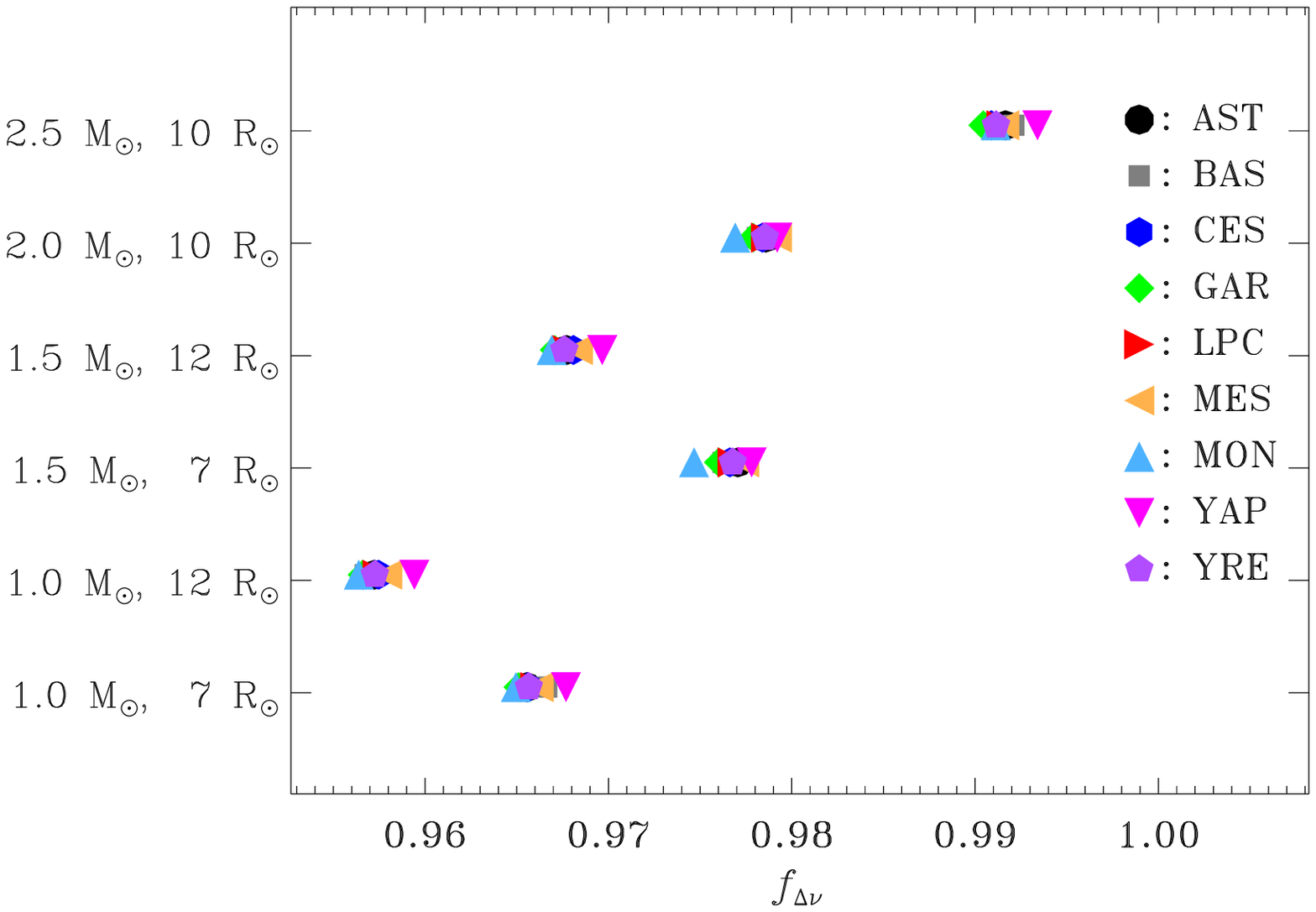}
\caption{
Correction factor $f_{\Delta \nu}$ for RGB-calibrated models
in the scaling relation for the
large frequency separation $\Delta \nu_{\rm fit}$ obtained from
fits to the radial-mode frequencies (cf.\ Eq.~\ref{eq:dnucalscale}).
The different codes are identified by the symbol shape and colour 
(cf.\ caption to Fig.~\ref{fig:diffdr0_solar}), with the addition of AST
(for {\tt ASTEC}).
}
\label{fig:dnusclteff}
\end{figure}

\begin{figure}[htpb]
\includegraphics[angle=0,scale=0.5]{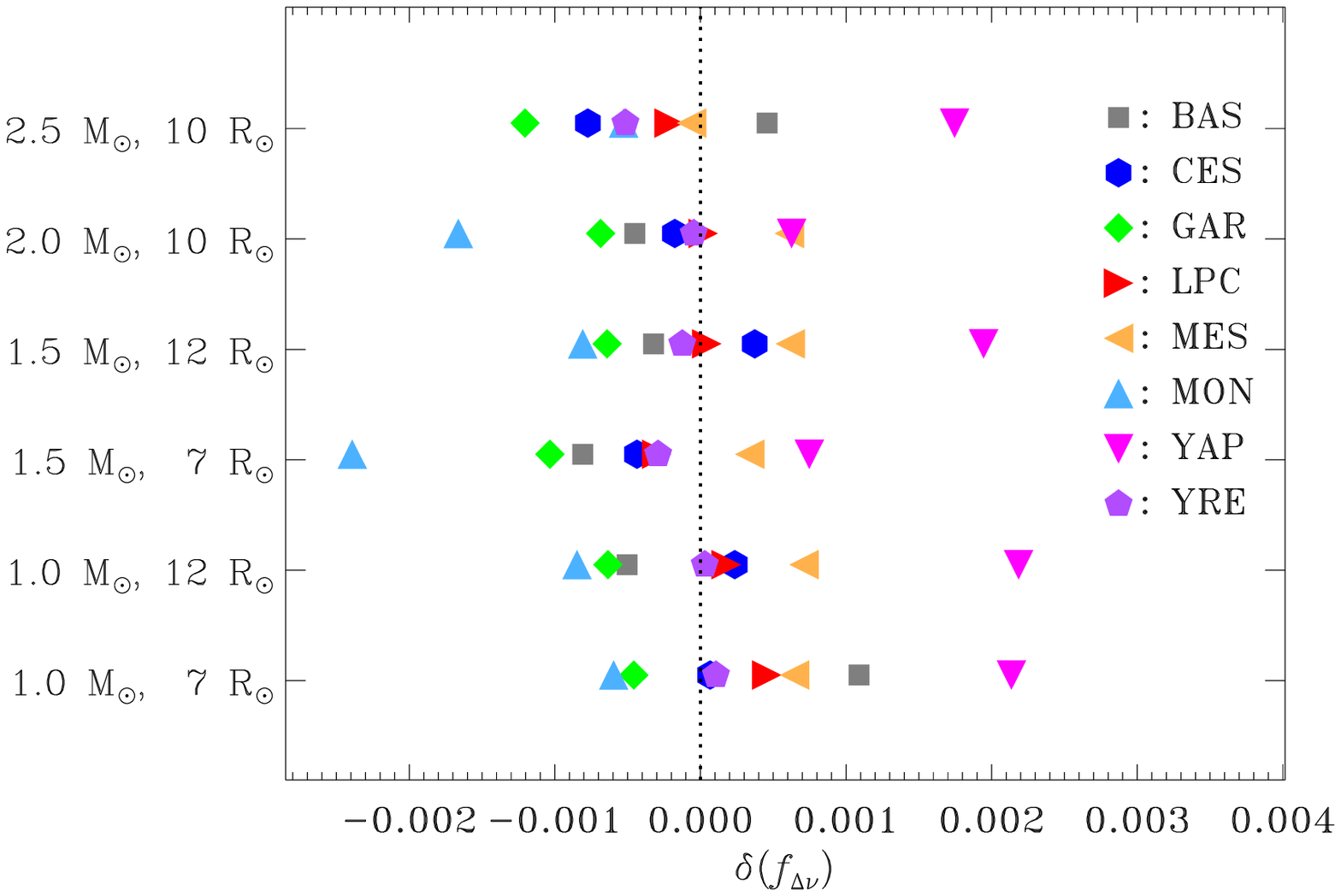}
\caption{
Differences relative to the {\tt ASTEC} model in the RGB-calibrated case,
in the sense (model) -- ({\tt ASTEC}),
in the correction factors $f_{\Delta \nu}$ (cf.\ Eq.~\ref{eq:dnucalscale})
relating the large frequency separation $\Delta \nu_{\rm fit}$ obtained from 
a fit to radial-mode frequencies and the value obtained
from homology scaling.
The different codes are identified by the symbol shape and colour 
(cf.\ caption to Fig.~\ref{fig:diffdr0_solar}).
\label{fig:diffdnusclteff}
}
\end{figure}

\begin{table*}[htpb]
\caption{Frequency $\nu_{\rm max}$, in $\muHz$, of maximum oscillation power
estimated from Eq.~(\ref{eq:numaxscale}) for RGB-calibrated models.}
\label{tab:numaxteff}
\centering
\begin{tabular}{r r c c c c c c c c c}
\hline\hline
$M/\msun$ & $R/\rsun$ & {\tt ASTEC} & {\tt BaSTI} & {\tt CESAM} & {\tt GARSTEC} & {\tt LPCODE} & {\tt MESA} & {\tt MONSTAR} & {\tt YAP} & {\tt YREC} \\
\hline
\smallskip
1.0 &  7.0 &  69.76 &  69.77 &  69.77 &  69.76 &  69.77 &  69.76 &  69.76 &  69.76 &  69.76\\
1.0 & 12.0 &  24.29 &  24.29 &  24.29 &  24.30 &  24.30 &  24.30 &  24.29 &  24.29 &  24.29\\
1.5 &  7.0 & 102.58 & 102.59 & 102.58 & 102.58 & 102.59 & 102.58 & 102.58 & 102.58 & 102.58\\
1.5 & 12.0 &  35.76 &  35.76 &  35.77 &  35.77 &  35.77 &  35.77 &  35.76 &  35.76 &  35.76\\
2.0 & 10.0 &  67.05 &  67.06 &  67.05 &  67.05 &  67.06 &  67.08 &  67.06 &  67.05 &  67.05\\
2.5 & 10.0 &  82.70 &  82.71 &  82.68 &  82.71 &  82.71 &  82.67 &  82.70 &  82.70 &  82.70\\
\hline
\end{tabular}
\end{table*}

\begin{table*}[htpb]
\caption{Asymptotic dipolar g-mode period spacings $\Delta \Pi_1$ in s 
(cf.\ Eq.~\ref{eq:dpi}) for RGB-calibrated models.
}
\label{tab:perspac}
\centering
\begin{tabular}{r r r r r r r r r r r}
\hline\hline
$M/\msun$ & $R/\rsun$ & {\tt ASTEC} & {\tt BaSTI} & {\tt CESAM} & {\tt GARSTEC} & {\tt LPCODE} & {\tt MESA} & {\tt MONSTAR} & {\tt YAP} & {\tt YREC} \\
\hline
\smallskip
1.0 &  7.0 &  72.12 &  71.76 &  72.27 &  72.22 &  72.74 &  72.88 &  73.08 &  76.03 &  73.12\\
1.0 & 12.0 &  58.36 &  57.83 &  58.18 &  58.31 &  58.50 &  58.74 &  58.91 &  60.31 &  58.95\\
1.5 &  7.0 &  69.91 &  69.64 &  70.13 &  70.06 &  70.45 &  70.86 &  70.97 &  74.43 &  71.03\\
1.5 & 12.0 &  57.31 &  56.79 &  57.26 &  57.23 &  57.34 &  57.59 &  57.75 &  59.52 &  57.88\\
2.0 & 10.0 &  78.72 &  78.57 &  77.73 &  76.82 &  78.12 &  77.39 &  79.30 &  81.91 &  79.39\\
2.5 & 10.0 & 123.90 & 123.45 & 120.69 & 120.83 & 122.43 & 122.61 & 124.47 & 125.03 & 123.53\\
\hline
\end{tabular}
\end{table*}

\begin{figure}[htpb]
\includegraphics[angle=0,scale=0.5]{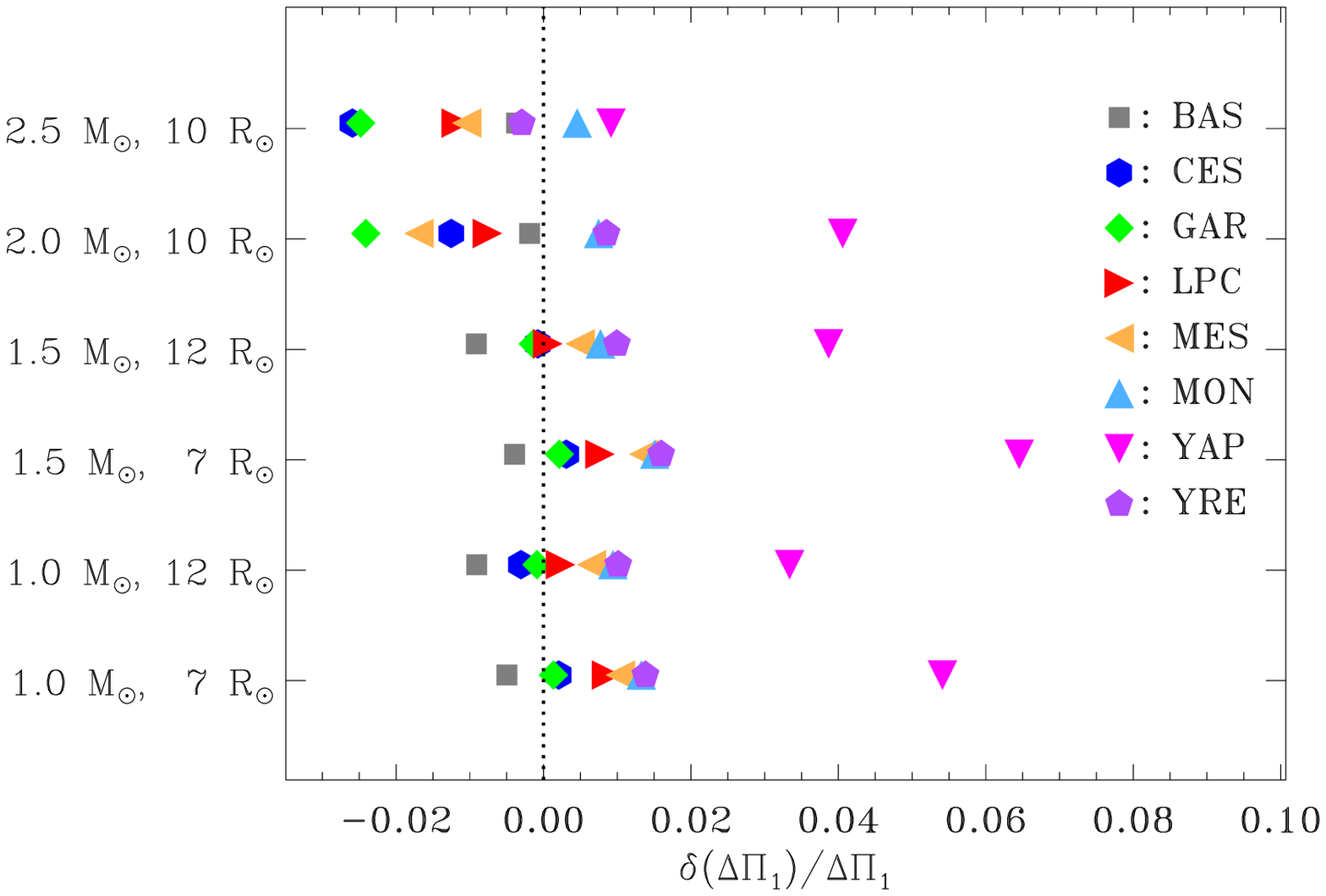}
\caption{
Relative differences for RGB-calibrated models
in the asymptotic period spacing $\Delta \Pi_1$,
compared with the {\tt ASTEC} results, in the sense (model) - ({\tt ASTEC});
the different codes are identified by the symbol shape and colour 
(cf.\ caption to Fig.~\ref{fig:diffdr0_solar}).
}
\label{fig:diffdpiteff}
\end{figure}

\clearpage

\vfill\eject

\section{Properties of the asymptotic large frequency separation}
\label{sec:dnuprop}

\begin{figure}[ht]
\includegraphics[angle=0,scale=0.5]{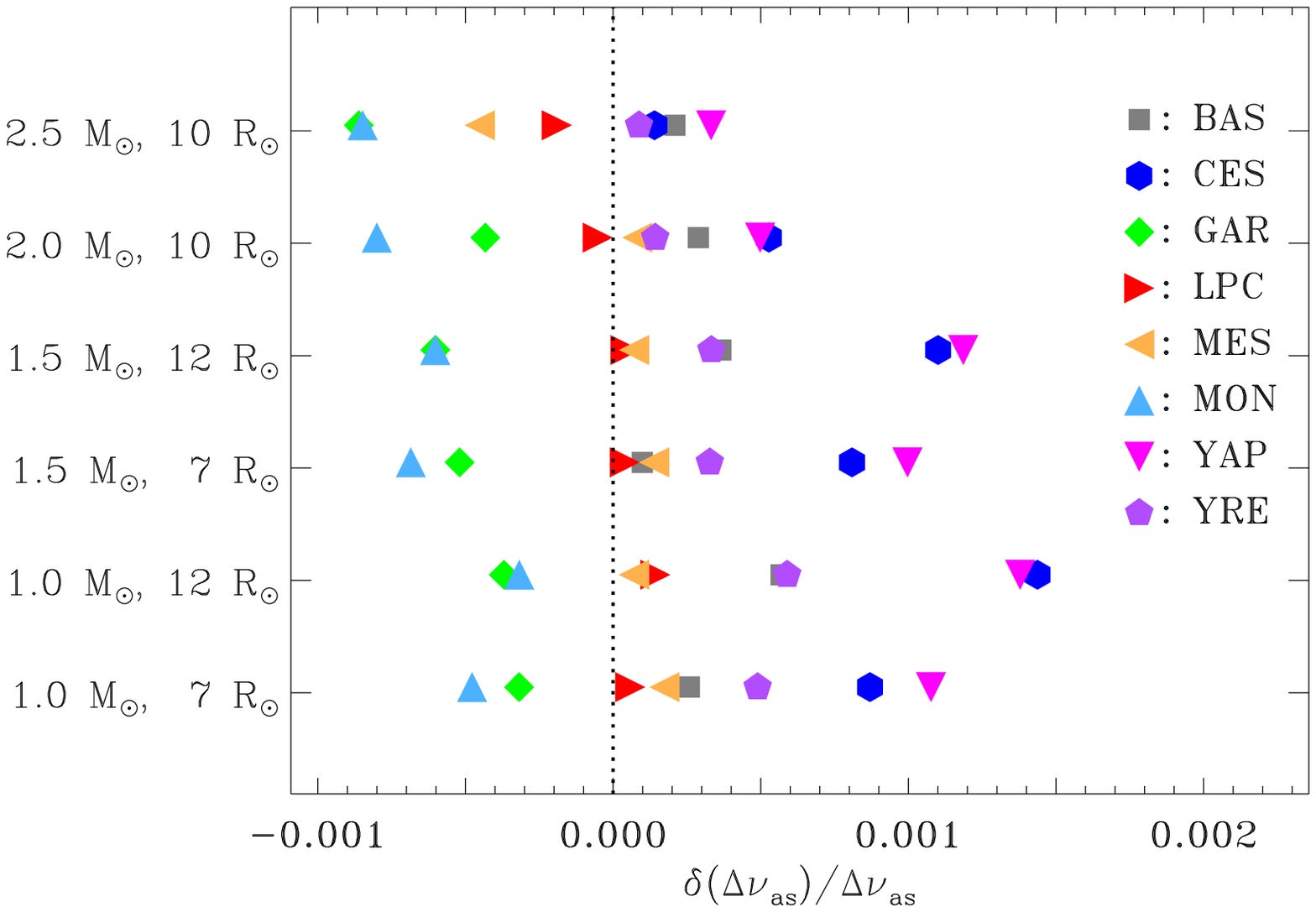}
\includegraphics[angle=0,scale=0.5]{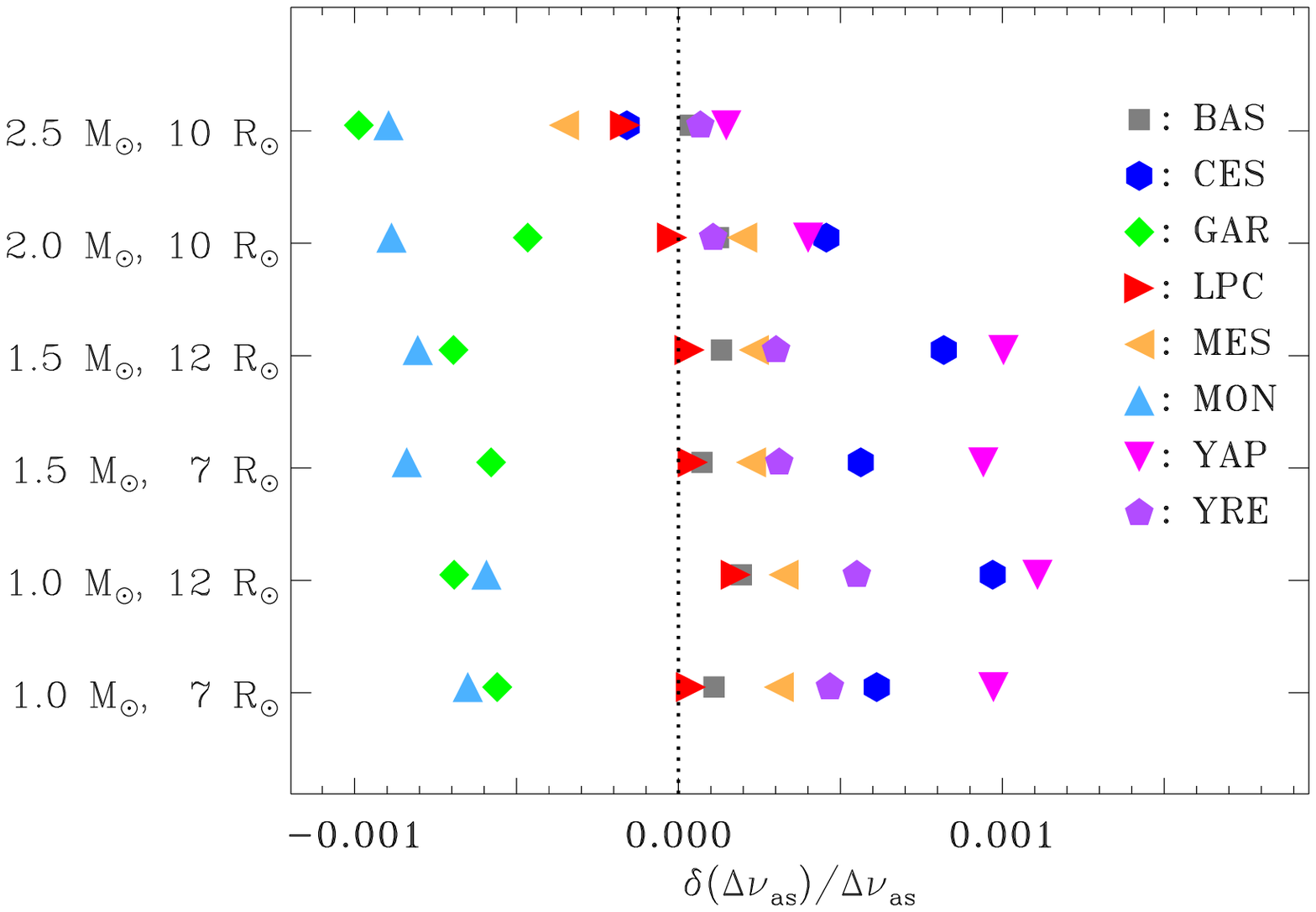}
\caption{
Relative differences in the asymptotic large frequency separation
$\Delta \nu_{\rm as}$,
compared with the {\tt ASTEC} results, in the sense (model) - ({\tt ASTEC});
the different codes are identified by the symbol shape and colour
and labelled by the abbreviated name of the code (see caption to
Fig.~\ref{fig:diffdr0_solar}).
The top panel shows results for solar-calibrated models and the bottom 
panel for the RGB-calibrated models.
}
\label{fig:diffdnusolar}
\end{figure}

\begin{table*}[ht]
\caption{Asymptotic acoustic-mode frequency separation $\Delta \nu_{\rm as}$
(cf.\ Eq.~\ref{eq:dnu}) in $\muHz$ for solar-calibrated models.
}
\label{tab:dnusolar}
\centering
\begin{tabular}{r r c c c c c c c c c}
\hline\hline
$M/\msun$ & $R/\rsun$ & {\tt ASTEC} & {\tt BaSTI} & {\tt CESAM} & {\tt GARSTEC} & {\tt LPCODE} & {\tt MESA} & {\tt MONSTAR} & {\tt YAP} & {\tt YREC} \\
\hline
\smallskip
1.0 &  7.0 &  7.792 &  7.794 &  7.799 &  7.790 &  7.793 &  7.794 &  7.789 &  7.801 &  7.796\\
1.0 & 12.0 &  3.507 &  3.509 &  3.512 &  3.505 &  3.507 &  3.507 &  3.506 &  3.512 &  3.509\\
1.5 &  7.0 &  9.504 &  9.505 &  9.512 &  9.500 &  9.504 &  9.505 &  9.497 &  9.513 &  9.507\\
1.5 & 12.0 &  4.263 &  4.264 &  4.267 &  4.260 &  4.263 &  4.263 &  4.260 &  4.268 &  4.264\\
2.0 & 10.0 &  6.432 &  6.434 &  6.435 &  6.430 &  6.432 &  6.433 &  6.427 &  6.435 &  6.433\\
2.5 & 10.0 &  7.227 &  7.229 &  7.228 &  7.221 &  7.226 &  7.224 &  7.221 &  7.229 &  7.228\\
\hline
\end{tabular}
\end{table*}
\begin{table*}[htpb]
\caption{Asymptotic acoustic-mode frequency separation $\Delta \nu_{\rm as}$
in $\muHz$ (cf.\ Eq.~\ref{eq:dnu}) for RGB-calibrated models.
}
\label{tab:dnuteff}
\centering
\begin{tabular}{r r c c c c c c c c c}
\hline\hline
$M/\msun$ & $R/\rsun$ & {\tt ASTEC} & {\tt BaSTI} & {\tt CESAM} & {\tt GARSTEC} & {\tt LPCODE} & {\tt MESA} & {\tt MONSTAR} & {\tt YAP} & {\tt YREC} \\
\hline
\smallskip
1.0 &  7.0 &  7.792 &  7.793 &  7.797 &  7.788 &  7.793 &  7.795 &  7.787 &  7.800 &  7.796\\
1.0 & 12.0 &  3.507 &  3.508 &  3.510 &  3.504 &  3.508 &  3.508 &  3.505 &  3.511 &  3.509\\
1.5 &  7.0 &  9.504 &  9.505 &  9.509 &  9.498 &  9.504 &  9.506 &  9.496 &  9.513 &  9.507\\
1.5 & 12.0 &  4.263 &  4.263 &  4.266 &  4.260 &  4.263 &  4.264 &  4.259 &  4.267 &  4.264\\
2.0 & 10.0 &  6.432 &  6.433 &  6.435 &  6.429 &  6.432 &  6.433 &  6.426 &  6.435 &  6.433\\
2.5 & 10.0 &  7.228 &  7.228 &  7.226 &  7.220 &  7.226 &  7.225 &  7.221 &  7.229 &  7.228\\
\hline
\end{tabular}
\end{table*}

Although we argue in Section~\ref{sec:genprop} that the 
asymptotic large frequency separation $\Delta \nu_{\rm as}$ does not 
provide an adequate accuracy for comparisons with observations
\citep[see also][]{Mosser2013}, 
it still represents the contribution from the bulk of the model to
the frequency separation.
Thus, it is of interest to compare $\Delta \nu_{\rm as}$ between the different
evolution codes.
Tables \ref{tab:dnusolar} and \ref{tab:dnuteff} show $\Delta \nu_{\rm as}$
for the solar-calibrated and RGB-calibrated models,
computed from Eq.~(\ref{eq:dnu}).
For simplicity we replace $R_*$ by $R_{\rm phot}$, the photospheric radius,
to avoid possible effects of differences in the models of the stellar
atmospheres.
The dominant variation of $\Delta \nu_{\rm as}$ with stellar properties 
follows the homology scaling,
$\Delta \nu \propto ( G M / R^3 )^{1/2}$ (see also Eq.~\ref{eq:dnuscale})
which, as discussed in Paper~I, is essentially fixed.
Thus, the variations between codes reflect more subtle differences in the
computed structure.
These variations are illustrated in Fig.~\ref{fig:diffdnusolar}, 
using the {\tt ASTEC} results as reference.
We note that the differences are substantially smaller than those found
for $\Delta \nu_{\rm fit}$ (cf.\ Fig.~\ref{fig:diffdnufrqsolar}).
This may be caused by differences in the structure of the
near-surface layers and atmospheres in the stellar models,
which would affect $\Delta \nu_{\rm as}$ less than the individual frequencies.
Also, these differences would have the strongest effect
on high-frequency modes, and hence may affect $\Delta \nu_{\rm fit}$ more
strongly than reflected in the root-mean-square frequency differences shown
in Fig.~\ref{fig:diffdr0_solar}.

\begin{figure}[htpb]
\includegraphics[angle=0,scale=0.5]{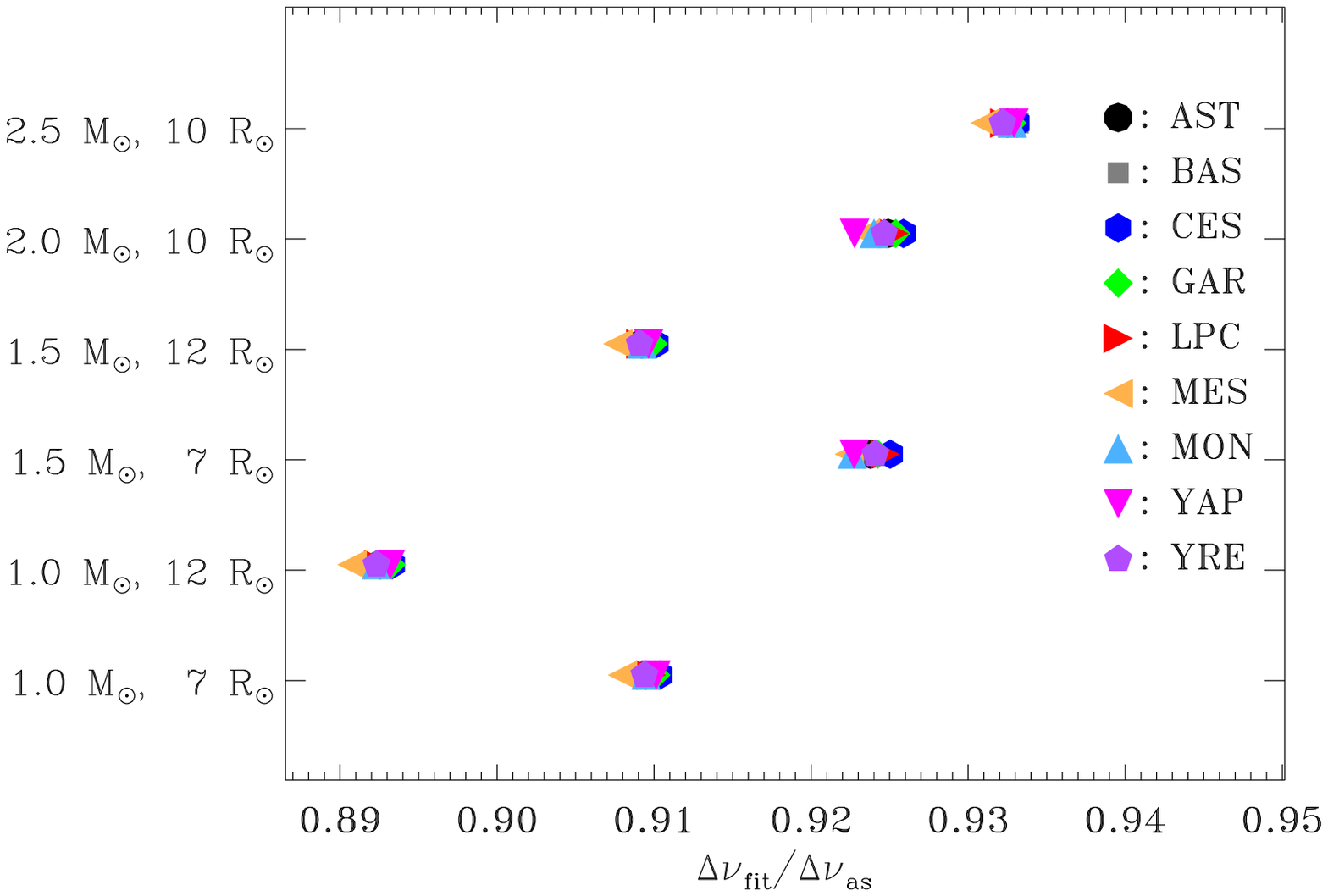}
\includegraphics[angle=0,scale=0.5]{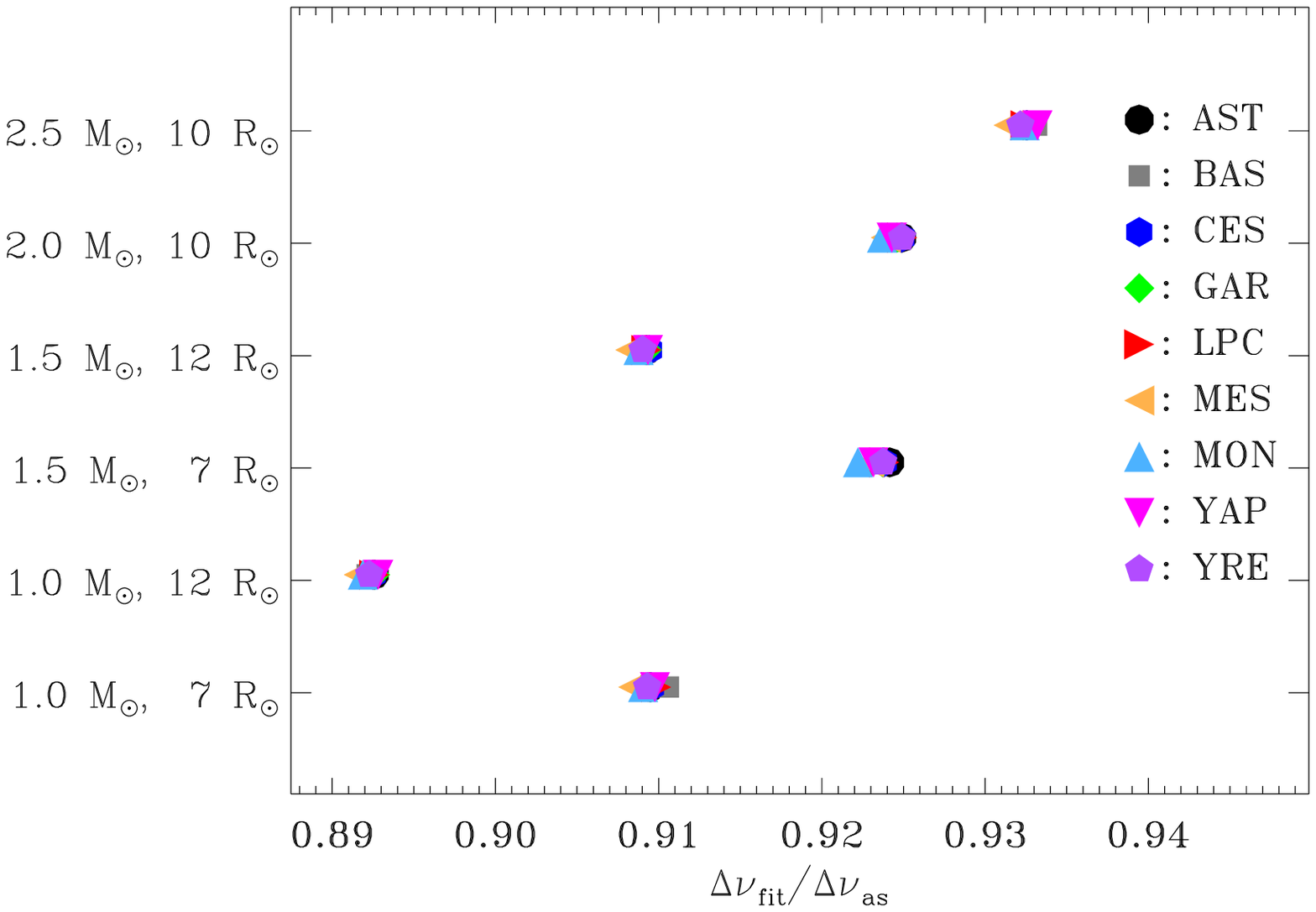}
\caption{
Ratios $\Delta \nu_{\rm fit}/\Delta \nu_{\rm as}$ between the 
large frequency spacing resulting from fit to the radial-mode frequencies
and the asymptotic values.
The symbols correspond to the different modelling codes, as defined in 
the caption to Fig.~\ref{fig:diffdr0_solar},
with the addition of AST ({\tt ASTEC}).
The upper panel shows results for the solar-calibrated models,
and the lower panel for the RGB-calibrated models.
}
\label{fig:dnufitas}
\end{figure}

The relation between the asymptotic value,
$\Delta \nu_{\rm as}$, of the large frequency separation 
(see Tables~\ref{tab:dnusolar} and \ref{tab:dnuteff})
and $\Delta \nu_{\rm fit}$ is of some interest.
Figure~Fig.~\ref{fig:dnufitas} shows their ratios.
It is evident that $\Delta \nu_{\rm as}$ substantially over-estimates
the actual value of $\Delta \nu$, no doubt to a large extent owing to the
choice of $R_{\rm phot}$ for the upper limit in the integral in 
Eq.~(\ref{eq:dnu}) rather than the location of the proper acoustic surface
(see the discussion below Eq. (\ref{eq:dnu}), and Section~\ref{sec:acoustic}).
This deserves further analysis.

\section{Problems with the MESA atmosphere models}
\label{sec:mesatm}

\begin{figure}[htpb]
\includegraphics[angle=0,scale=0.5]{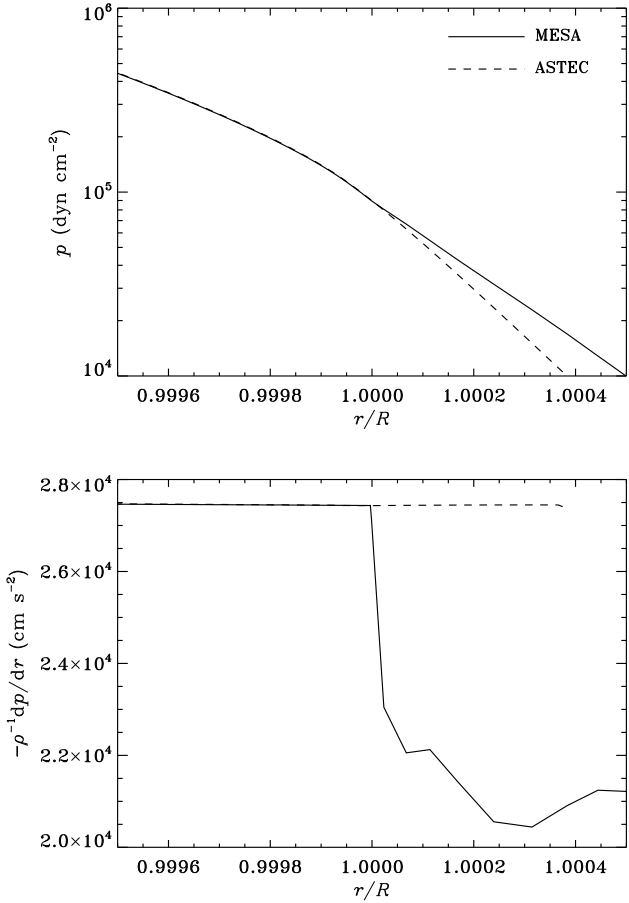}
\caption{
Top panel: pressure $p$ in $(1 \msun, 1 \rsun)$ models computed with
{\tt MESA} (solid line) and {\tt ASTEC} (dashed line).
Bottom panel: the pressure gradient divided by density for these models.
}
\label{fig:mespres}
\end{figure}

\begin{figure}[htpb]
\includegraphics[angle=0,scale=0.5]{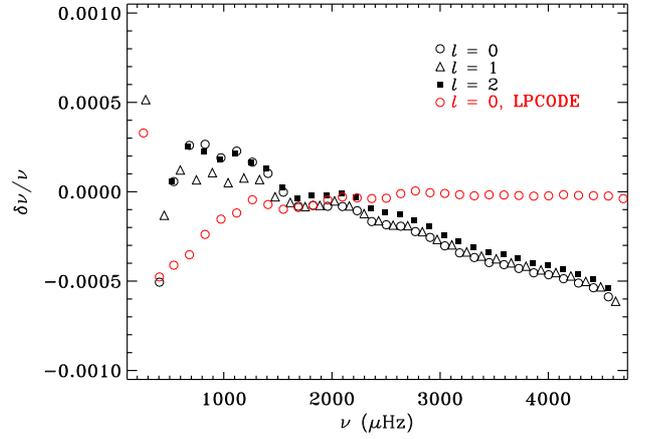}
\caption{
Relative frequency differences for the $(1 \msun, 1 \rsun)$ 
solar-calibrated case between {\tt MESA} and {\tt ASTEC},
in the sense ({\tt MESA}) -- ({\tt ASTEC}),
for $l = 0$ (open circles), $l = 1$ (open triangles) and $l = 2$
(filled squares) in black.
The differences are evaluated at fixed radial order.
The red circles show corresponding results for radial modes in 
the {\tt LPCODE} model.
\label{fig:mespresfreq}
}
\end{figure}

In Fig.~\ref{fig:dfmesm15r7} we found an increase in magnitude 
in the frequency differences between the {\tt MESA} and the {\tt ASTEC}
models at high frequency.
This reflects what appears to be a general problem with the structure 
of the atmospheric models in version 6950 of {\tt MESA} used in the
present comparison.
We illustrate this by considering the $(1 \msun, 1 \rsun)$ model
in the solar-calibrated comparison track.
Very similar effects are found in the other MESA cases considered.

Figure~\ref{fig:mespres} compares the pressure and its derivative in the
{\tt MESA} and {\tt ASTEC} models.
The top panel shows substantial differences between the atmospheric pressures
in the two models, whereas they are essentially in agreement below the 
photosphere.
The difference between the models is even more dramatic in the bottom panel: 
this shows $- \rho^{-1} \dd p / \dd r$, which according to 
the equation of hydrostatic equilibrium 
should be equal to the gravitational acceleration $g$ and
hence essentially constant in the outermost parts of the model.
This is satisfied in the {\tt ASTEC} model but not in the {\tt MESA} model.
The effect on the computed frequencies is shown in Fig.~\ref{fig:mespresfreq},
compared also with the results for the corresponding {\tt LPCODE} model.
For the {\tt MESA} model there are clearly significant differences, particularly
at high frequency, as expected for model differences confined to the outermost
layers; no such differences are found in the case of the {\tt LPCODE}
(although there are differences at low frequency, which reflect structure
differences deeper in the model).

For the present comparisons these problems with the {\tt MESA} models have
a relatively minor effect, compared with the more substantial differences
found for various other aspects of the structure.
However, they would affect the comparison between observations and the 
{\tt MESA} models, and more generally it has clearly been desirable to correct
these problems with such a convenient and widely used code.
We note that they have been resolved in MESA since revision 11877.

\section{Asteroseismic effects of thermodynamic properties}
\label{sec:asteos}

\begin{figure}[htpb]
\includegraphics[angle=0,scale=0.5]{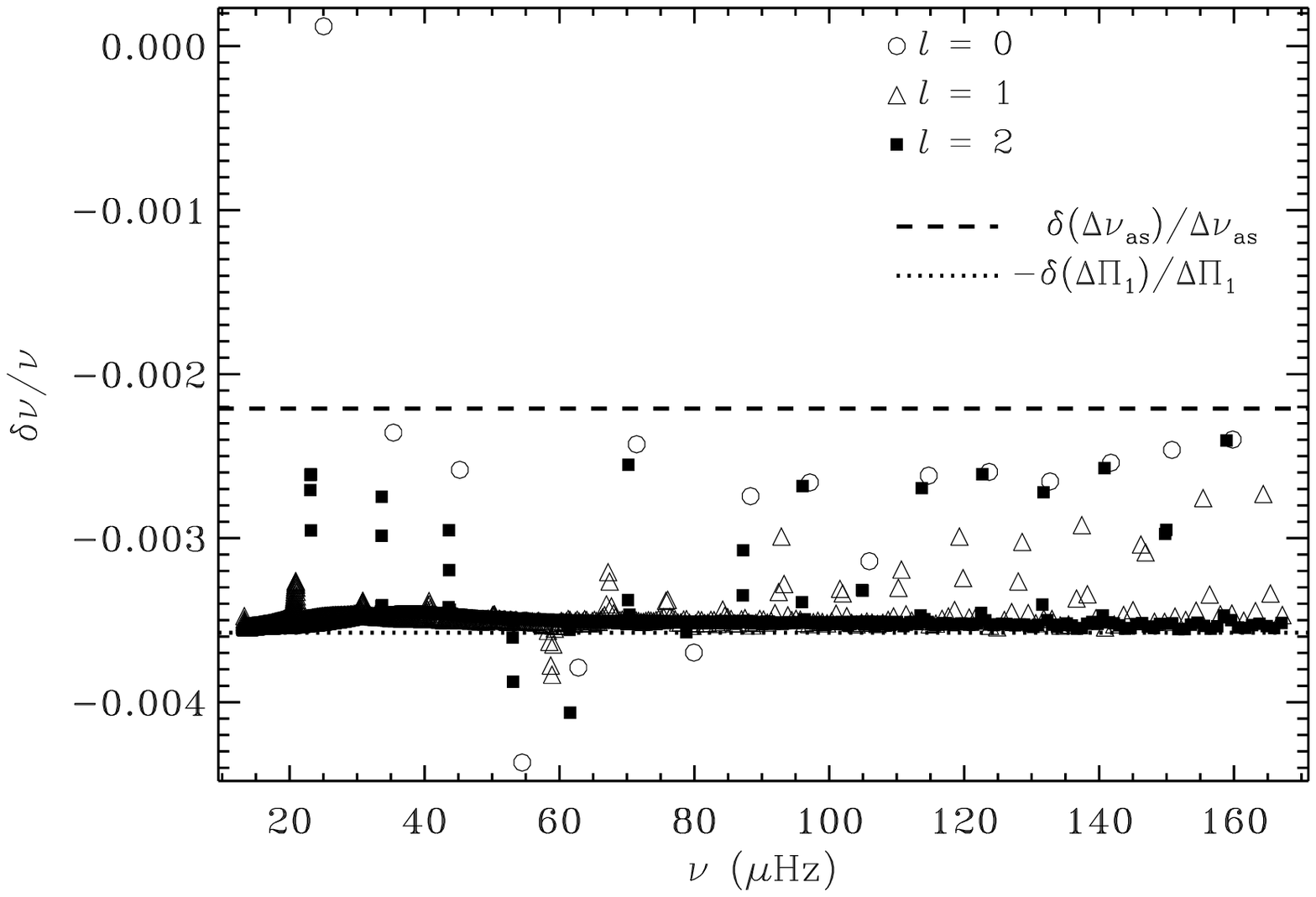}
\includegraphics[angle=0,scale=0.5]{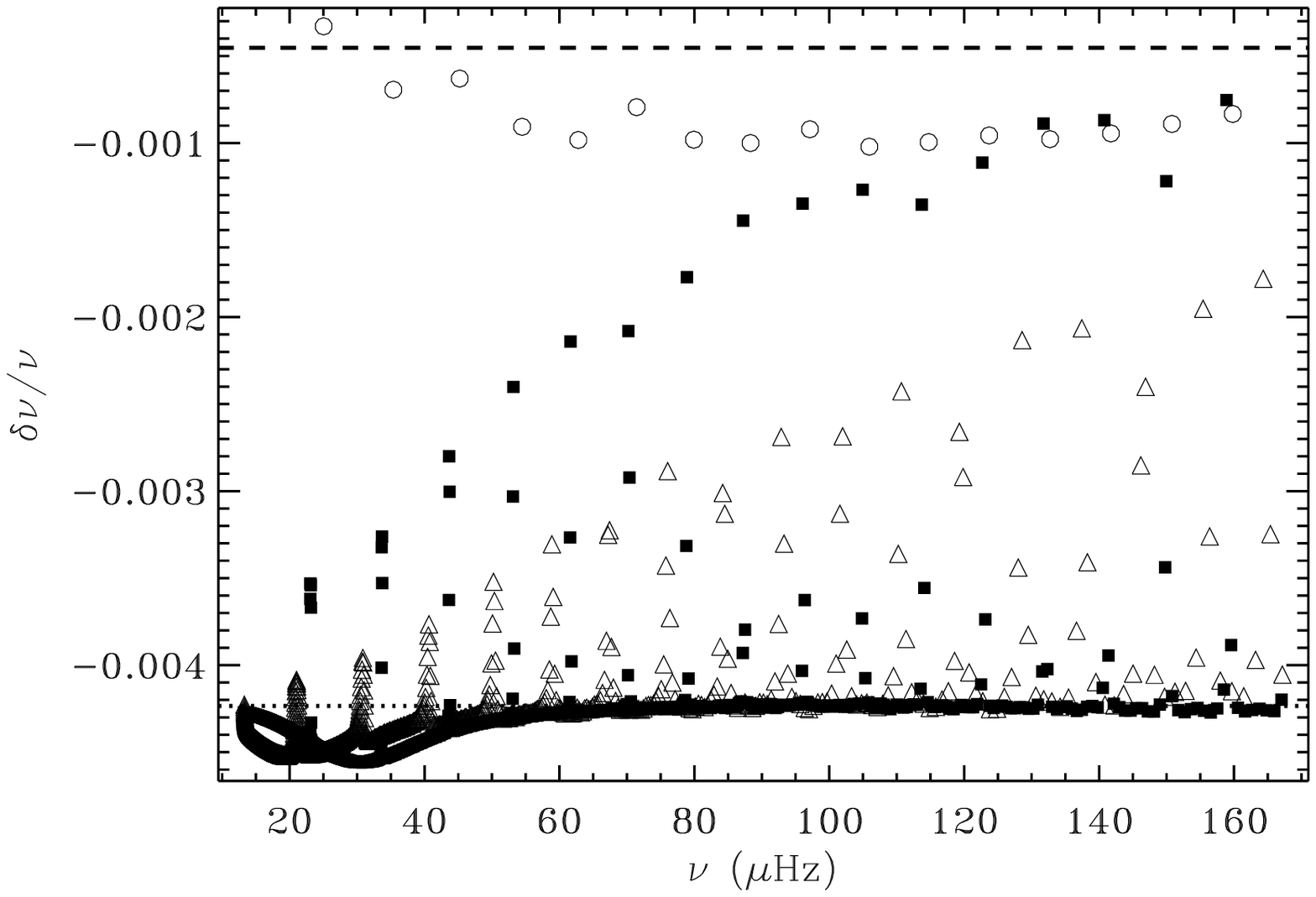}
\caption{
Relative differences in computed frequencies for the {\tt GARSTEC}
$1.5 \msun, 7 \rsun$ solar-calibrated models, 
compared with the {\tt ASTEC} results, in the sense
({\tt GARSTEC}) - ({\tt ASTEC}).
The top panel shows results for the original {\tt GARSTEC} model,
and the bottom panel used the revised model, with updated 
treatment of the equation of state.
The differences are evaluated at fixed radial order.
}
\label{fig:dfgarm15r7}
\end{figure}

The original results for the {\tt GARSTEC} models showed rather 
substantial differences, relative to the {\tt ASTEC} reference,
in the acoustic-mode properties.
These arose from a separate treatment
in the version of {\tt GARSTEC} used then of the low-temperature region 
in the implementation of the OPAL equation of state.
This has been updated in the results shown in the main part of this paper.
However, since the results provide insight into the sensitivity of the
frequencies to the model structure it is of interest to discuss them
in some detail.

Frequency differences between the original {\tt GARSTEC} and the
{\tt ASTEC} models
for the $1.5 \msun, 7 \rsun$ solar-calibrated case 
are shown in the top panel of Fig.~\ref{fig:dfgarm15r7}.
Compared with the general trends in the average radial-mode frequency
differences shown in Fig.~\ref{fig:diffdr0_solar},
there are substantial differences in
the radial-mode frequencies and in the asymptotic frequency spacing 
and a significant discrepancy between the differences in 
the asymptotic and actual radial-mode frequencies.
The differences in the asymptotic
period spacing and consequently in the g-dominated mode frequencies
are comparatively small; they are coincidentally similar to the radial-mode
frequency differences.

\begin{figure}[htpb]
\includegraphics[angle=0,scale=0.5]{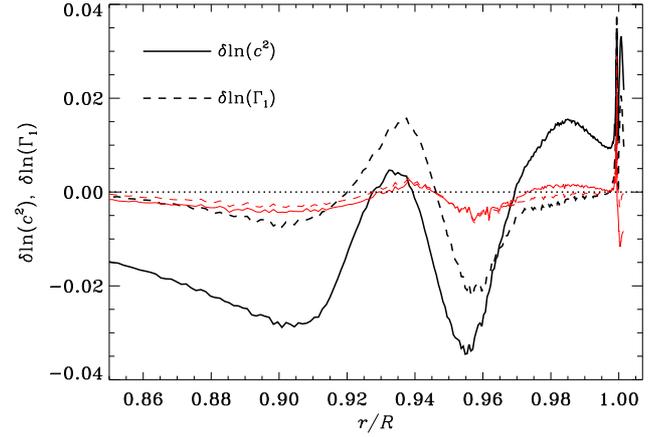}
\caption{
Logarithmic differences between the original {\tt GARSTEC} and
{\tt ASTEC} $1.5 \msun, 7 \rsun$ solar-calibrated models,
in the sense ({\tt GARSTEC}) - ({\tt ASTEC}), in the outer layers of the model.
The solid line shows the difference in squared sound speed $c^2$ and the dashed 
line the difference in adiabatic exponent $\Gamma_1$.
For comparison, the thinner red lines show the corresponding differences
between the revised {\tt GARSTEC} and the {\tt ASTEC} models.
}
\label{fig:diffcsq}
\end{figure}

These differences in acoustic behaviour between the original {\tt GARSTEC}
and {\tt ASTEC} models
are directly related to differences in the structure of the outer layers
of the models.
Figure~\ref{fig:diffcsq} shows the logarithmic differences in squared
sound speed $c^2 = \Gamma_1 p/\rho$ and $\Gamma_1$.
It is evident that much of the sound-speed difference comes from the
difference in $\Gamma_1$, in the region of helium ionisation.
This is the result of significant differences between the models 
in the treatment of the equation of state in these regions.
As shown by the red curves, these differences have been very substantially
reduced by the revision of the {\tt GARSTEC} models.

With the revised {\tt GARSTEC} equation of state the differences in
acoustic behaviour between {\tt GARSTEC} and {\tt ASTEC} are very
small, as illustrated by the bottom panel of Fig.~\ref{fig:dfgarm15r7}.

\section{Asteroseismic effects of the convective-core size}
\label{sec:astccore}

\begin{table}[htpb]
\caption{Asymptotic dipolar g-mode period spacings $\Delta \Pi_1$ in s 
	(cf.\ Eq.~\ref{eq:dpi}) for the solar-calibrated {\tt ASTEC} and
	the original and corrected {\tt LPCODE} models.
}
\label{tab:perspacsolarold}
\centering
\begin{tabular}{r r r r r}
\hline\hline
$M/\msun$ & $R/\rsun$ & {\tt ASTEC} & {\tt LPCODE} & {\tt LPCODE} \\
          &           &             & (original)   & (corrected) \\
\hline
\smallskip
2.0 & 10.0 &  78.72 & 73.37 &  78.14 \\
2.5 & 10.0 & 123.62 & 117.11 & 122.32 \\
\hline
\end{tabular}
\end{table}

\begin{figure}[htpb]
\includegraphics[angle=0,scale=0.5]{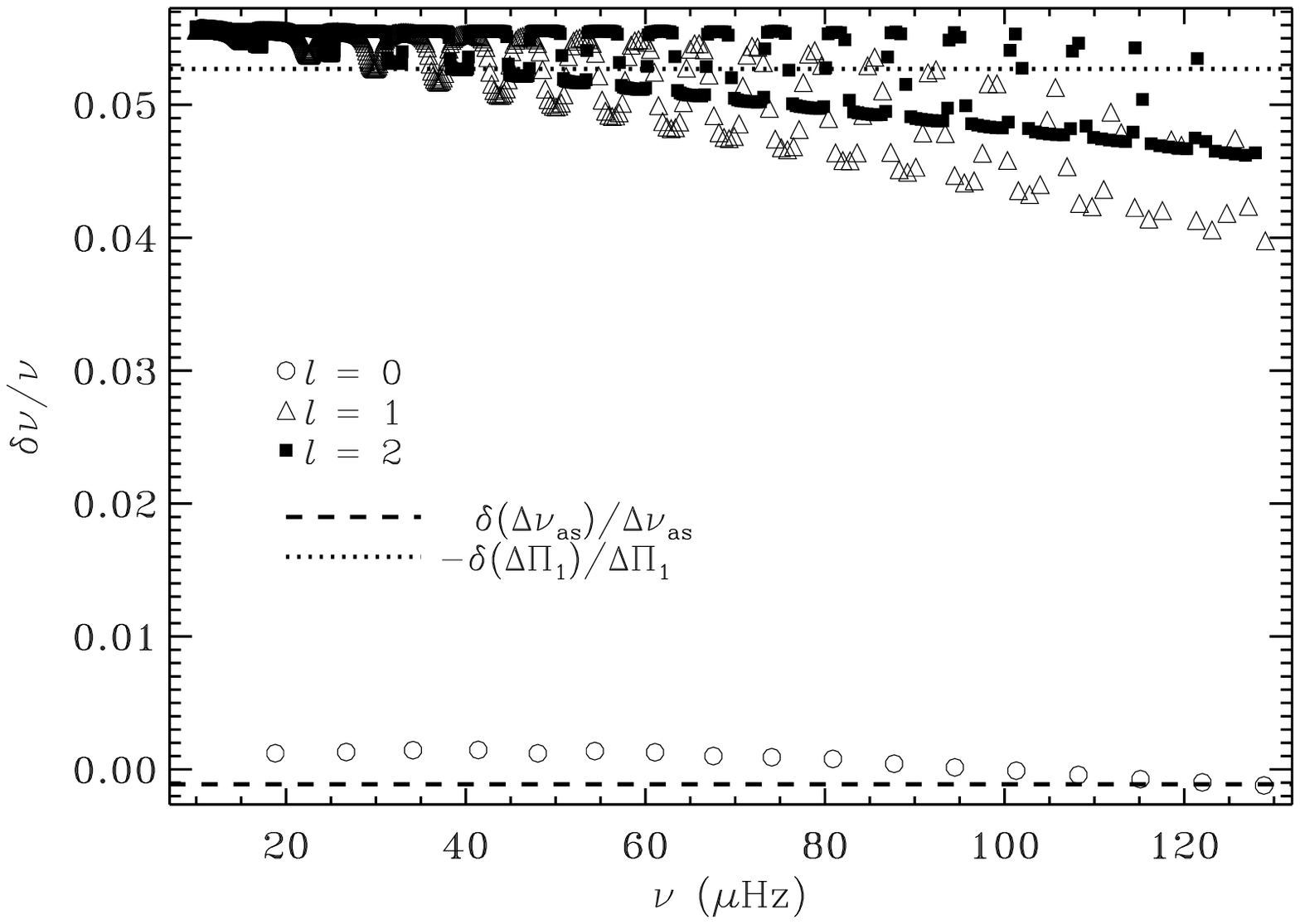}
\includegraphics[angle=0,scale=0.5]{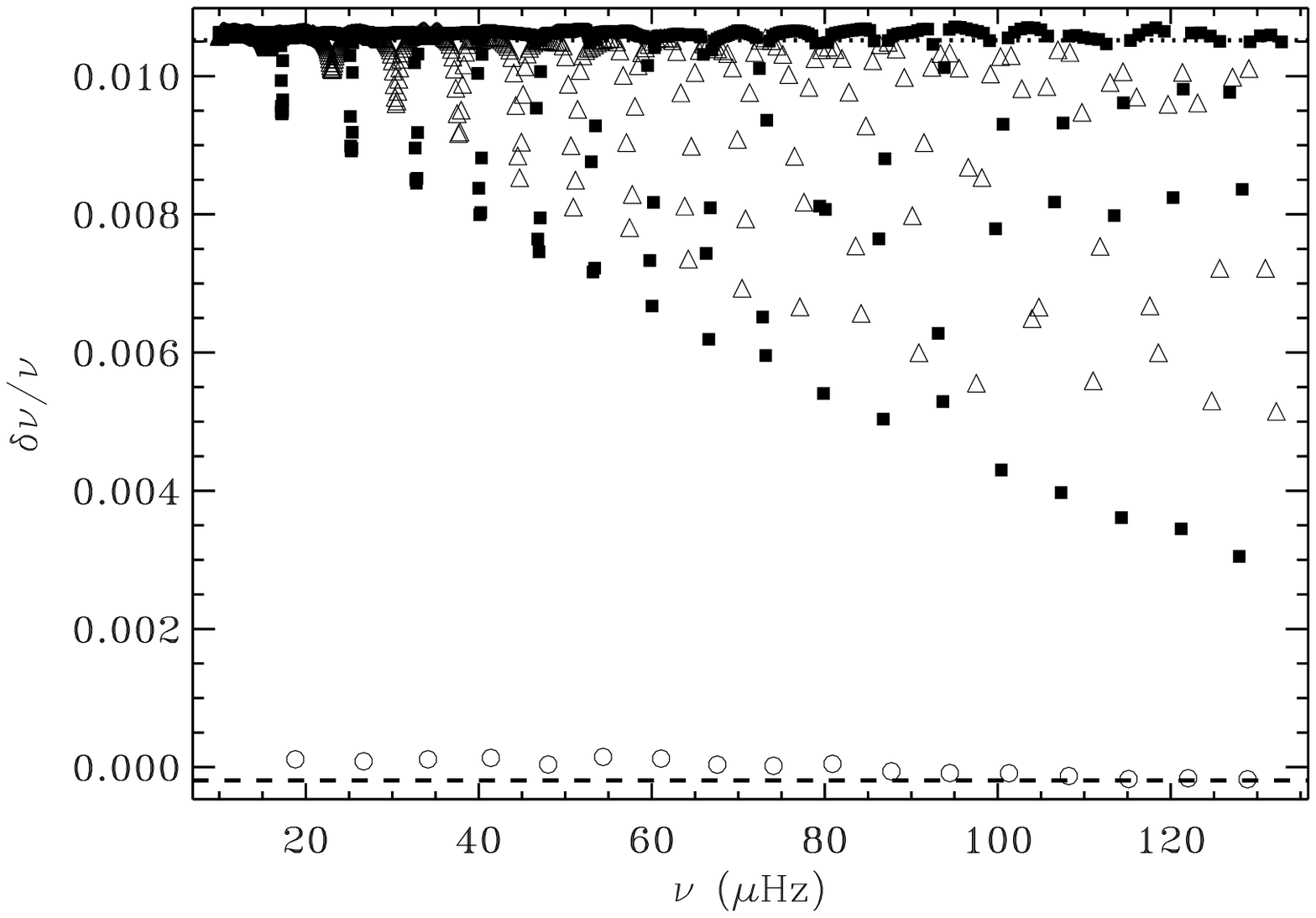}
\caption{
Relative differences in computed frequencies for the {\tt LPCODE}
$2.5 \msun, 10 \rsun$ solar-calibrated model,
compared with the {\tt ASTEC} results,
'n the sense ({\tt LPCODE}) - ({\tt ASTEC}).
The top panel shows the original {\tt LPCODE} model,
while the bottom panel is for the revised model.
}
\label{fig:dflpcm25r10old}
\end{figure}

\begin{figure}[ht]
\includegraphics[angle=0,scale=0.5]{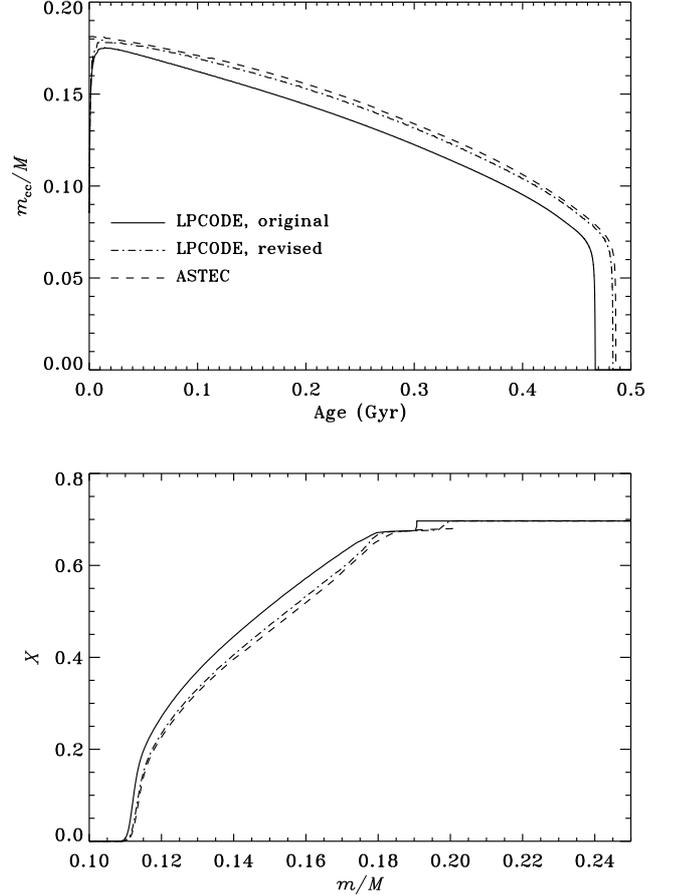}
\caption{
Top panel: variation with age in the fractional mass of the convective core,
in $2.5 \msun$ solar-calibrated models.
Bottom panel: resulting profiles of the hydrogen-abundance $X$ in the
$10 \rsun$ red-giant model.
The solid and dot-dashed lines show the original and revised {\tt LPCODE}
models,
and the dashed lines show the corresponding {\tt ASTEC} model.
}
\label{fig:xproflpcm25r10}
\end{figure}

The original comparisons found substantial differences in g-mode frequencies 
between the $2.5 \msun, 10 \rsun$ {\tt LPCODE} and {\tt ASTEC} models,
as illustrated in the top panel of Fig.~\ref{fig:dflpcm25r10old}. 
Here there is excellent agreement for the radial-mode frequencies, while the
g-dominated modes, and the asymptotic period spacing, show differences of
around 5\,\%.
The pattern of differences is qualitatively similar 
to Fig.~\ref{fig:dfmesm15r7}, with smaller differences for the p-dominated 
modes.
Similar effects were found in the $2.0 \msun, 10 \rsun$ case, 
as illustrated by $\Delta \Pi_1$ in Table~\ref{tab:perspacsolarold}.
These differences were caused by differences in the H profile in the region
located between the H-burning shell and the H-discontinuity left by
the first dredge-up episode
(see the bottom panel in Fig.~\ref{fig:xproflpcm25r10}).
Differences in the chemical profile can be traced back to a
smaller receding convective core during main-sequence evolution
due to an underestimation of the radiative opacities in the core. 
The latter was caused by the fact that the {\tt OPAL} routines
\citep{Iglesi1993} were using the Type II set of opacity tables%
\footnote{see {\tt https://opalopacity.llnl.gov/existing.html}}
as soon as C and O were transformed into  N by the CNO cycles.
As the C and O decrement was not balanced by the N-enhancement 
in the opacity tables, this led to a slight underestimation of
the Rosseland opacity of the core. 
This has now been corrected in {\tt LPCODE}, as shown in Section~\ref{sec:res}
and the bottom panel of Fig.~\ref{fig:dflpcm25r10old}.
However, the results provide an illustrative example of the effect
on red-giant frequencies of changes to the main-sequence convective core
and hence deserves a more detailed analysis.
Here we focus on the solar-calibrated case; the RGB-calibrated case is
very similar.

Relevant properties of the evolution and structure are presented in 
Fig.~\ref{fig:xproflpcm25r10} (see also Paper~I).
The top panel shows the evolution in the fractional mass of the convective core,
which defines the hydrogen profile at the end of the main sequence.
It is evident that the convective core is significantly smaller, 
and the main-sequence phase correspondingly shorter, in the original
{\tt LPCODE} evolution, whereas the corrected evolution is very similar to
the {\tt ASTEC} case.
In the $10 \rsun$ model this is reflected in a slightly smaller helium-rich
region in the original {\tt LPCODE} model.
Correcting the opacity increases the size of the convective core to
close to but still slightly smaller than the {\tt ASTEC} model,
resulting in the much smaller frequency differences shown in
the bottom panel of Fig.~\ref{fig:dflpcm25r10old}.

\begin{figure}[ht]
\includegraphics[angle=0,scale=0.5]{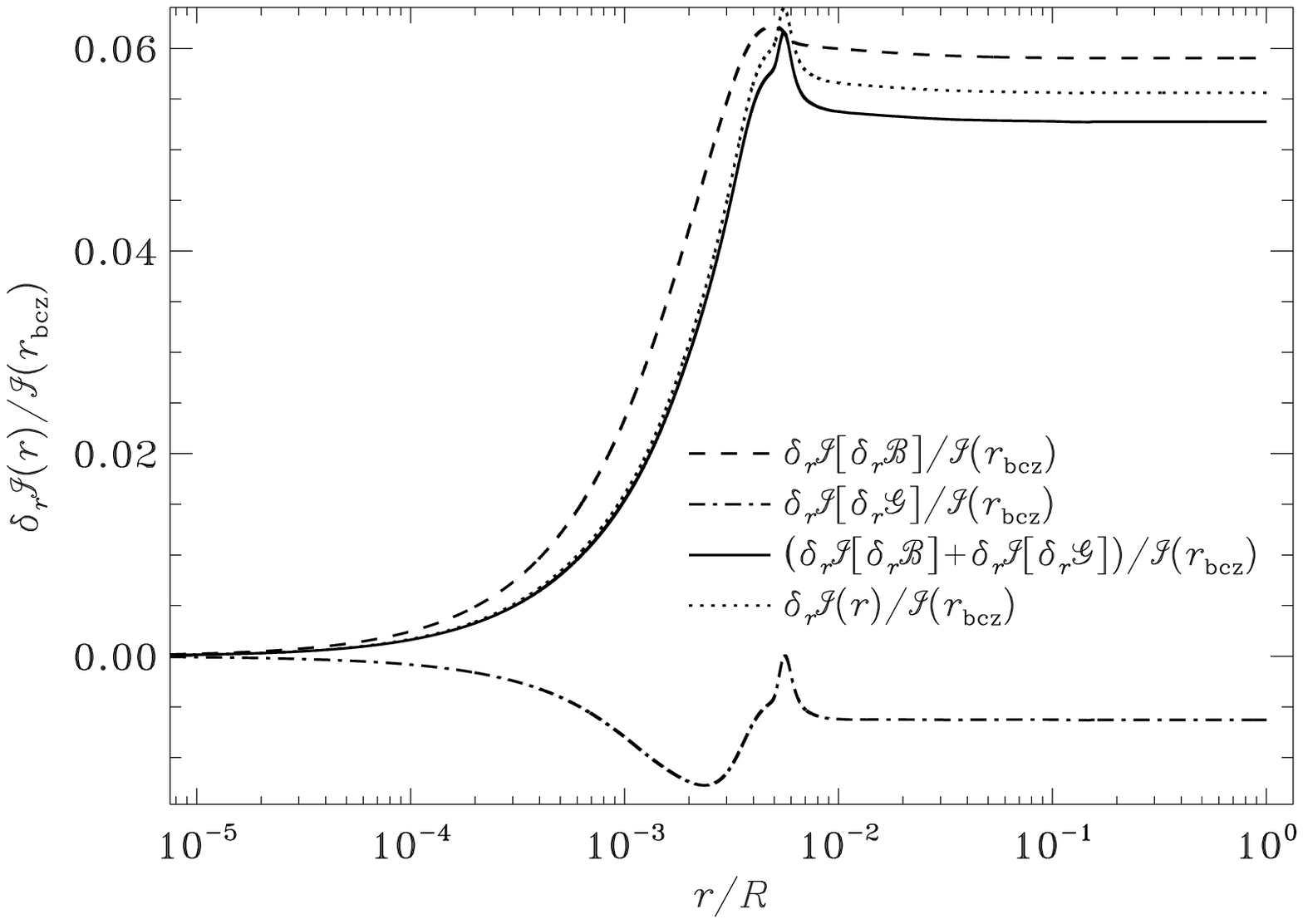}
\caption{
Differences in the partial integrals for differences in $\Pi_0$
(cf. Eq.~\ref{eq:dcbuoyint})
between the original {\tt LPCODE} $2.5 \msun, 10 \rsun$ solar-calibrated model 
and the {\tt ASTEC} model, in the sense ({\tt LPCODE}) -- ({\tt ASTEC}).
The dashed and dash-dotted lines show the contributions
$\delta_r \CI[\delta_r \CB]$ and $\delta_r \CI[\delta_r \CG]$ from
$\delta_r \CB$ and $\delta_r \CG$, respectively, and the solid line
shows their sum.
For comparison, the dotted line shows the relative difference in $\CI$.
}
\label{fig:dalpcm25r10}
\end{figure}

To investigate how the differences in structure affect
the asymptotic period spacing, we express $\Pi_0$ (cf.\ Eq.~\ref{eq:dpi}) as
\begin{equation}
\Pi_0 = 2 \pi^2 \CI(\rbcz)^{-1} \; ,
\label{eq:dpii}
\end{equation}
where $\rbcz$ is the radius at the base of the convective envelope, and
\begin{equation}
\CI(r) = \int_0^{r} N {\dd r \over r} \; .
\label{eq:buoyint}
\end{equation}
Also, using the equation of hydrostatic support we introduce
\begin{equation}
N^2 = \CB \CG \; ,
\label{eq:buoy1}
\end{equation}
separating $N^2$ in a dynamical and a thermodynamic part, with
\begin{equation}
\CB = {g^2 \rho \over p} \; , \qquad
\CG = \left( {1 \over \Gamma} - {1 \over \Gamma_1} \right)  \; ,
\label{eq:buoycont}
\end{equation}
where
\begin{equation}
{1 \over \Gamma} = {\dd \ln \rho \over \dd \ln p} \; .
\end{equation}
From Eq.~(\ref{eq:dpii}) it follows that
\begin{equation}
\delta \ln \Pi_0 \simeq - {\delta_r \CI(\rbcz) \over \CI(\rbcz)} \; ,
\label{eq:dbuoyint}
\end{equation}
neglecting the small contribution from the difference in $\rbcz$
between the models;
here, from Eqs.~(\ref{eq:buoyint}), (\ref{eq:buoy1}) and (\ref{eq:buoycont}),
\begin{equation}
\delta_r \CI(r) \simeq {1 \over 2} \int_{0}^r N (\delta_r \ln \CB + 
\delta_r \ln \CG ) {\dd r \over r} \; ,
\label{eq:dcbuoyint}
\end{equation}
where $\delta_r$ denotes the difference at fixed fractional radius.
The result of the analysis is shown in Fig.~\ref{fig:dalpcm25r10}.
It is clear that Eq.~(\ref{eq:dcbuoyint}) provides a reasonable approximation
to the difference in $\CI$, which is dominated by the contribution
$\delta_r \CI [\delta_r \CB]$ from $\delta_r \ln \CB$.

\begin{figure}[ht]
\includegraphics[angle=0,scale=0.5]{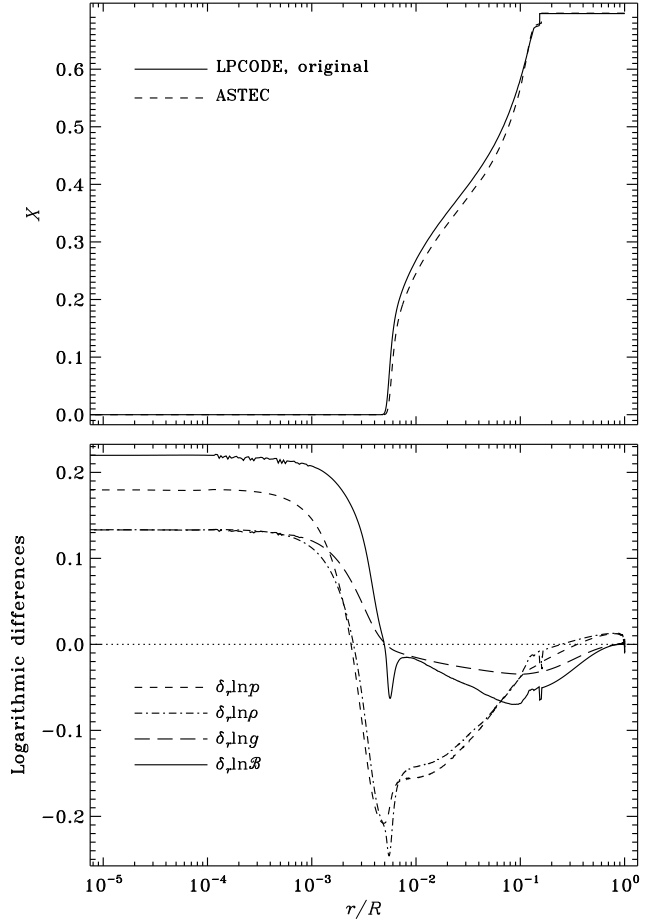}
\caption{
Top: hydrogen abundance in the {\tt ASTEC} (dashed) and 
the original {\tt LPCODE} (solid) $2.5 \msun, 10 \rsun$ solar-calibrated models.
Bottom: differences $\delta_r \ln p$ (dashed), 
$\delta_r \ln \rho$ (dot-dashed), $\delta_r \ln g$ (long dashed) and 
$\delta_r \ln \CB$ (solid) between the {\tt ASTEC} 
and the original {\tt LPCODE} $2.5 \msun, 10 \rsun$ models,
in the sense ({\tt LPCODE}) -- ({\tt ASTEC}).
}
\label{fig:dfglpcm25r10}
\end{figure}

To delve deeper into the origin of these differences, 
Fig.~\ref{fig:dfglpcm25r10} shows the hydrogen abundance 
in the {\tt LPCODE}
and {\tt ASTEC} models, as well as the logarithmic differences
between the models in $p$, $\rho$, $g$ (i.e.\ the mass) and $\CB$.
The differences are predominantly in and just above the core of the model,
probably related to the difference in the hydrogen profile.
The change in the partial integral $\CI(r)$ (cf.\ Fig.~\ref{fig:dalpcm25r10})
is dominated by the core, where the {\tt LPCODE} model has a higher central
condensation and the larger gravitational acceleration dominates the
difference in $\CB$.
It is interesting that the asymptotic period spacing and the mixed-mode
frequencies so clearly reflect the relatively subtle difference in the
hydrogen profile.

\end{appendix}
\clearpage
%
%
%

\end{document}